\documentclass[pdflatex,sn-apa]{sn-jnl}

\usepackage{pgfplots} 
\usepackage{verbatim} 
\usepackage{tcolorbox}
\usepackage{dirtree}
\usepackage{multirow}
\usepackage{longtable}
 \usepackage{xspace}
 \usepackage{xfrac}
\usepackage{soul}
\usepgfplotslibrary{groupplots} %
\usepackage{graphicx}%
\usepackage{multirow}%
\usepackage{amsmath,amssymb,amsfonts}

\usepackage{amsthm}%
\usepackage{mathrsfs}%
\usepackage[title]{appendix}%
\usepackage{xcolor}%
\usepackage{colortbl}
\usepackage{textcomp}%
\usepackage{manyfoot}%
\usepackage{booktabs}%
\usepackage{algorithm}%
\usepackage{algorithmicx}%
\usepackage{algpseudocode}%
\usepackage{listings}%
\usepackage{float}   
\usepackage{tikz}
\usetikzlibrary{shapes.geometric, arrows,positioning}

 \definecolor{lightblue}{RGB}{180, 180, 220}  
\definecolor{pastel}{RGB}{151, 188, 203} 
\definecolor{darkerblue}{RGB}{0, 100, 130}

\newcommand{\OpenAI}{OpenAI\xspace}

\usepackage[nohyperlinks,nolist]{acronym}

\usepackage{tabularx}
\usepackage{soul}

\begin{document}

\begin{acronym}
    \acro{Code LLM}{Code \ac{LLM}}
    \acro{LLM}{Large Language Model}
\end{acronym}

\title[Article Title]{ Navigating Pitfalls: Evaluating LLMs in Machine Learning Programming Education }


\author*[1]{\fnm{Smitha} \sur{Kumar}}\email{smitha.kumar@hw.ac.uk}

\author[2]{\fnm{Michael} \sur{A. Lones}}\email{m.lones@hw.ac.uk}

\author[2]{\fnm{Manuel} \sur{ Maarek}}\email{m.maarek@hw.ac.uk}

\author[1]{\fnm{Hind} \sur{ Zantout}}\email{h.zantout@hw.ac.uk}

\affil[1]{\orgdiv{
School of Mathematical and Computer Sciences}, \orgname{Heriot-Watt University}, \orgaddress\country{United Arab Emirates}}
\affil[2]{\orgdiv{
School of Mathematical and Computer Sciences}, \orgname{Heriot-Watt University}, \orgaddress\country{United Kingdom}}


\abstract{ The rapid advancement of Large Language Models (LLMs) has opened new avenues in education. This study examines the use of  LLMs  in supporting learning in  machine learning education; 
in particular,   
it focuses on the ability of LLMs to identify common errors of practice (pitfalls) in machine learning code, and their ability to provide feedback that can guide learning. Using a portfolio of code samples, we consider four different LLMs: one closed model and three open models. Whilst the most basic pitfalls are readily identified by all models, many common pitfalls are not. They particularly struggle to identify pitfalls in the early stages of the ML pipeline, especially those which can lead to information leaks, a major source of failure within applied ML projects. They also exhibit limited success at identifying pitfalls around model selection, which is a concept that students often struggle with when first transitioning from theory to practice. This questions the use of current LLMs to support machine learning education, and also raises important questions about their use by novice practitioners. Nevertheless, when LLMs successfully identify pitfalls in code, they do provide feedback that includes advice on how to proceed, emphasising their potential role in guiding learners. We also compare the capability of closed and open LLM models, and find that the gap is relatively small given the large difference in model sizes. This presents an opportunity to deploy, and potentially customise, smaller more efficient LLM models within education, avoiding risks around cost and data sharing associated with commercial models.

 } 

\keywords{large language models, automated feedback generation, machine learning pitfalls, computer science education }



\maketitle

\section{Introduction}\label{sec1} 
\acp{LLM}, and especially specialised code-generating LLMs such as \OpenAI Codex~\citep{openAI}, GitHub Copilot~\citep{githubGitHubCopilot}, StarCoder~\citep{huggingfaceBigcodestarcoderHugging}, and CodeQwen~\citep{huggingfaceQwenCodeQwen157BHugging} are revolutionizing  core software engineering related tasks like code generation~\citep{austin2021programsynthesislargelanguage}, code interpretation~\citep{10.1145/3545945.3569785}, testcase generation~\citep{liu2024llmpoweredtestcasegeneration}, code review~\citep{10.1145/3545945.3569759}, code debugging~\citep{wermelinger2023using} and program repair~\citep{10.1145/3618305.3623587}.  Researchers are actively investigating the implications of these \acp{LLM} on the educational landscape, especially in the context of programming education ~\citep{10.1145/3623762.3633499,10.1145/3587103.3594206}. 
While these models offer a range of potential benefits to educators and students, the majority of research on LLMs within an education context has concentrated on introductory programming courses (CS1 and CS2). 
As the complex nature  of programming is overwhelming for many student programmers \citep{goldman2008identifying, denny2012all,ettles2018common}, various techniques---both Machine Learning (ML) and non-ML based---have been proposed in the past to generate clear, concise, personalized formative feedback \citep{keuning2018systematic}. Recent research indicates that \acp{LLM} outperform these earlier approaches \citep{xia2023automated} and have significant potential as a digital teaching assistant to support learning \citep{10.1145/3649217.3653574}. 

There has been comparatively little focus on the use of LLMs within ML education, and no studies (at the time of writing) looking at their specific use within ML code generation and feedback within an educational context. This is despite the increasing importance of ML  within education. For instance, universities and online learning platforms report a significant rise in ML  enrollments, and introductory ML concepts are increasingly being taught prior to tertiary level. Importantly, ML programming differs from traditional imperative programming, with a focus on systems-level construction rather than low-level syntax and use of control structures. Consequently, the type of programming errors committed by students in ML are also different \citep{skripchuk2022identifying,zimmermann2023common}. ML also differs from traditional programming to the extent to which learning continues beyond formal education and into practice. The prevalence of  errors made during this transition to practice
is currently a major concern in ML \citet{Lones2024}, further motivating the need for the present study into how LLMs can support learning in ML programming, and whether their current capabilities are sufficient in this regard.

The study is guided by the following research questions:
 
\begin{description}
    \item [RQ1] -  How reliable are LLMs at identifying common machine learning pitfalls in programs written for university-level assignments?
    \item [RQ2] - How useful is the feedback generated by LLMs in response to machine learning code samples that contain common pitfalls?
\end{description}

This study also considers the use of open LLM models within education. 
Proprietary (or closed) LLMs are widely used; however, they present challenges in terms of financial constraints and ethical considerations around student data privacy \citep{leinonen2024llmitationsincerestformdata}. Open models offer cost-effective solutions with greater customization options to create digital teaching assistants for feedback generation \citep{KASNECI2023102274}. Presently, there is little understanding of how open models compare to closed models in terms of their ability to provide feedback on ML code.

We make the following key contributions:
\begin{itemize}
    \item Measuring, for the first time, the ability of code LLMs to identify common pitfalls within machine learning code, finding that all the models have limited effectiveness, with recall rates of less than 50\%.
    \item  Showing that LLMs perform better at detecting pitfalls relevant to early stages of ML education, and perform poorly on those more typical of real world practice. We also note a bias towards correctly identifying pitfalls later in the ML pipeline.
    \item Comparing the abilities of open and closed LLM models, and finding the differences to be relatively small given the large differences in model size. This suggests that freely-deployable open models may be a viable alternative to commercial models.
    \item Finding that, when they identify errors, all the LLMs typically provide feedback to learners that includes information about how to proceed, which is valuable within an educational context.
    \item Publicly sharing a repository\footnote{\url{https://github.com/ssk705/ML_Pitfalls}} of our dataset and supporting scripts in order to support replication and further studies.
\end{itemize}

The paper is organised as follows. Section~\ref{sec:background} reviews background in ML education, code LLMs, and their use in programming education. Section~\ref{sec:methodology} presents our methodology, including the pitfalls, code samples and LLMs that we use, and the process for prompting, collecting and evaluating responses. Section~\ref{sec:results} presents results and discussion, organised around the two research questions. Limitations and conclusions are presented in Sections~\ref{sec:limitations} and~\ref{sec:conclusions} respectively.

\section {Background} \label{sec:background}
This discussion examines education and practice in machine learning, the emergence of LLMs, and their growing significance in reshaping programming education.  

\subsection{Machine learning education and practice}
ML is a prominent sub-field of artificial intelligence which focuses on developing data-driven models for classification, prediction, and regression tasks. ML systems have become ubiquitous across various domains, from the mundane to the critical. This increasing prevalence of ML is reflected by a growing demand and emphasis on ML education, both within computer science programmes and more broadly. Effective ML 
courses involve conveying both the underlying concepts of ML and how they can be integrated and applied practically. Key underlying concepts include data cleaning techniques, feature selection and transformation, model training, tuning mechanisms, and evaluation processes and metrics.

Early challenges in ML education include developing adequate mathematical and programming skills~\citep{cinca2025they}, in order to understand and implement underlying concepts. However, in interviews with educators, \citet{sulmont2019can} found that the most challenging elements lie later in the curriculum when students need to assemble components into systems capable of solving real world problems. This shift emphasises systems-level thinking as students transition from 
understanding the theory to implementing ML models in practice, and this is mirrored in the use of open-ended ML tasks within final year undergraduate and masters-level programmes. Two published studies~\citep{skripchuk2022identifying,zimmermann2023common} identified common mistakes made by students in such tasks, particularly the lack of systematic hyperparameter tuning, the misuse of test data, and inappropriate model selection or component usage. However, mistakes were observed across all stages of the ML pipeline.

Errors in ML practice are also common beyond formal education. Partly, this is due to the high demand for ML coupled with the relative recency of the field, resulting in most ML practitioners being in the early stages of their careers and lacking deep practical experience. This is particularly evident in the academic and scientific context, where numerous studies have identified and reported errors made in the design or implementation of published ML processes~\citep{liao2021we,Lones2024,ref1}. In many cases, these errors have led to false conclusions and poor replicability~\citep{KAPOOR2023100804}. While specific applications of ML in industry may be less transparent to scrutiny, it has been widely reported that a 
considerable proportion of ML/AI projects in industry also fail~\citep{cooper2024ai}, and failures in the implementation phase appear to be a major contributing factor~\citep{ermakova2021beyond}. Furthermore, issues related to AI education and literacy have also been flagged as real challenges within broader industrial product development~\citep{nahar2023meta}.

\subsection {Large language models for code}
Language models are computational models that can process and generate natural language. The field has undergone a   transformation from early rule-based systems such as ELIZA \citep{weizenbaum1966eliza} to neural-based multimodal transformers such as GPT-4 \citep{zhao2024surveylargelanguagemodels}. Transformers utilize a self-attention mechanism to capture contextual relationships within text to generate coherent and contextually appropriate outputs with remarkable proficiency \citep{naveed2024comprehensiveoverviewlargelanguage}. The impressive language skills of contemporary LLMs can be attributed to several factors, including their size (number of trainable parameters), volume of training data, and context window size (their maximum prompt size) \citep{burtsev2024working}. LLMs are trained on massive amounts of text, first through the semi-supervised task of predicting the next token, and then through various forms of instruction-fine-tuning so that they can be used within question-answering services such as ChatGPT.

Whilst most contemporary LLMs often appear to understand code, code LLMs are explicitly trained on large collections of code samples, in addition to non-code data such as natural language. They are designed to solve programming tasks such as code generation, code summarization, code translation and bug detection.  Contemporary code LLMs are based on the same transformer backbones as general-purpose LLMs and, just like these, are trained to respond to natural language prompts. The Code LLM landscape has grown significantly in recent times with a large number of open and closed models such as Code Llama, GitHub Copilot, DeepSeek Coder, and Qwen~2.5-Coder \citep{jiang2024surveylargelanguagemodels}.
 
\subsection {Code LLMs in programming education}
Code LLMs are having a profound impact on the field of programming education. One benefit lies in the ability for students to get instant feedback on their code, something that is known to have a positive impact on student engagement and performance \citep{10.1145/2843043.2843070}. Computer science researchers have extensively explored various approaches to create formative and summative feedback \citep{price2019comparison,gulwani2018automated,yi2017feasibility}. 
According to Narciss \citep{narciss2008feedback}, “feedback refers to all post-response information which informs the learner on his/her actual state of learning or performance in order to regulate the further process of learning” and the various types of feedback include  knowledge of performance (KP), knowledge of result or response (KR), and knowledge of the correct response (KCR). Elaborated feedback types such as knowledge about task constraints (KTC), knowledge about concepts (KC), knowledge about mistakes (KM), knowledge about how to proceed (KH), and knowledge about meta-cognition (KMC) offer greater guidance to learners.

Understanding of the challenges faced by novice programmers is essential for delivering meaningful feedback. Research studies indicate that novice programmers often struggle to master key concepts such as algorithm effectiveness, program design, and object-oriented concepts \citep{goldman2008identifying} \citep{ettles2018common}. To address this, there is a growing interest among researchers to explore the potential of code LLMs within an introductory programming context  \citep{sobania2023analysisautomaticbugfixing, zhang2022repairingbugspythonassignments,phung2023generatinghighprecisionfeedbackprogramming,xia2023automated,10.1145/3643795.3648380}. An example of a system based around this approach is CodeTutor\citep{10.1145/3657604.3662036}, which has demonstrated positive results on student performance, although it was not found to contribute significantly to critical thinking skills. The study also emphazised the importance of  well-crafted prompts for effective responses. 

Despite their potential benefits for student learning, there are growing concerns about over-reliance on LLM-based code-generation models \citep{Prather_2023}. There have also been many documented cases of LLMs producing incorrect code \citep{tambon2024bugs}, and a study of responses to student programmers’ help requests indicated that misleading feedback was also common \citep{hellas2023exploring}. Nevertheless, studies have found that code LLMs can be a valuable education tool if steps are taken to mitigate the risks related to incorrect responses, student overdependence, and academic misconduct \citep{10.1145/3626252.3630958}.

This study focuses on ML code. Code written to solve ML tasks is written in the same programming languages as general code, but differs significantly in its structure and style. It typically takes the form of high-level scripts which tie together calls to various lower-level library functions, and in this respect presents a more linear and more system-oriented approach than traditional forms of programming. Nevertheless, standard code LLMs can be used for generating ML code and there is evidence that they can generate useful ML code~\citep{cheng2023gpt}, and  their widespread adoption for this purpose~\citep{nejjar2025llms}. Tu et al.~\citep{tu2023should} explore the potential applications of LLMs within data science education, which encompasses ML.

With regard to the earlier discussion on errors in ML code and their prevalence, a significant risk associated with using LLMs in this manner is that there is a high probability that the LLM’s training data contains numerous instances of incorrect ML code. This poses a particular risk of LLMs generating incorrect code and providing erroneous feedback on code. However, to the best of our knowledge, there has been limited research on comprehending the specific risks associated with using LLMs for ML code-related tasks, whether in educational contexts or more broadly. This study seeks to address this gap.


\section{Methodology} 

\begin{figure}[tb]
    \centering
    \includegraphics[width=0.9\linewidth]{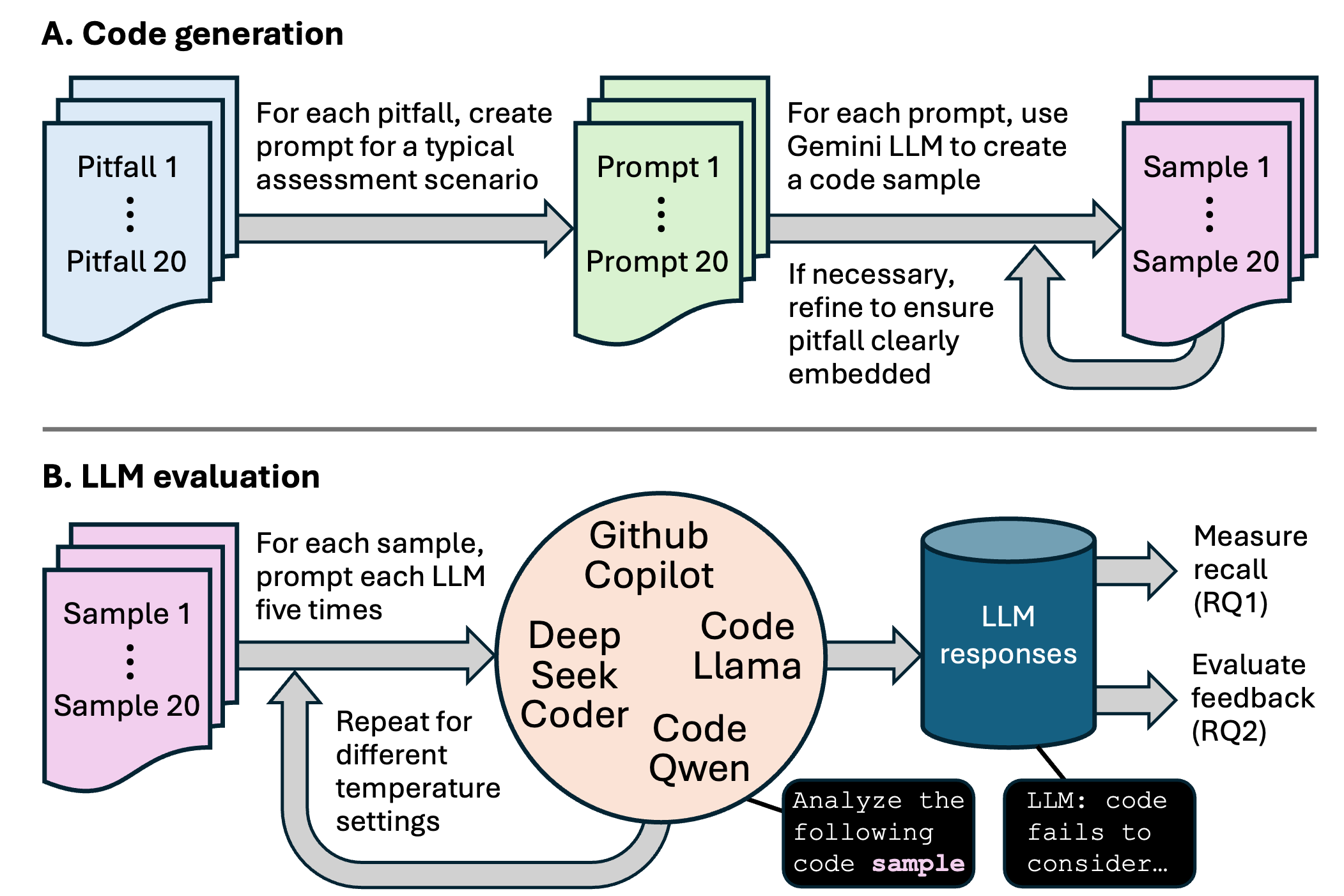}
    \caption{Overview of methodology, showing (A) generation of machine learning code samples used to assess LLMs, and (B) evaluation of LLM responses when prompted with these code samples.}
    \label{fig:methodology}
\end{figure}

\label{sec:methodology}
This section describes our methodology, which is conveyed in Figure~\ref{fig:methodology}. Section~\ref{sec:methodology:pitfalls} outlines the machine learning pitfalls considered in this work. Section~\ref{sec:methodology:samples} describes the process used to generate relevant code samples that exhibit these pitfalls. Section~\ref{sec:methodology:llms} lists the LLMs which we are evaluating. Section~\ref{sec:methodology:prompt} shows the prompt design. Section~\ref{sec:methodology:evaluation} describes the collection and evaluation of responses to this prompt.

\subsection {Catalogue of machine learning pitfalls}  \label{sec:methodology:pitfalls}
Many types of errors can occur in ML pipelines. This study focuses on those  which can be identified through static analysis of code alone, i.e.\ through function calls, their arguments and the ordering of statements. The specific choice of pitfalls, summarised in Table~\ref{tab:Mlpitfalls}, is informed by two recent guides to common errors made within ML practice \citep{Lones2024,ref1} and also include those observed in student code in previous studies \citep{skripchuk2022identifying,zimmermann2023common}. The pitfalls cover the various stages of the ML pipeline, from preprocessing and splitting of data through to the evaluation and comparison of trained models, and all relate to errors that practitioners routinely make when they first apply ML methods. Some of these pitfalls are relatively easy to identify through code; for example, those, such as PF\_03--05, that relate to whether preprocessing is done before or after splitting data. Others are more challenging, particularly those which require the LLM to take into account characteristics of the dataset, e.g.\  PF\_06, which requires the LLM to recognise that a dataset is being used which likely contains multiple samples per subject.

\begin{table} [tb!]
    \centering
    \begin{tabularx}{\hsize}{@{}lX@{}}
    \toprule
        ML Pitfall & Description \\
        \midrule
        PF\_01 No test & Model trained and tested on the same data. This results in a misleading measure of model generalisability.
        \textit{California, LR}\\
        PF\_02 Reuse test & Multiple models are evaluated on the same test data. This can lead to over-confident measures of performance.
        \textit{California, LR}\\
        PF\_03 Scale early & Scaling parameters are determined using the whole dataset. This allows information about test data to leak into training.
        \textit{Iris, RF}\\
        PF\_04 Select early & Feature selection is done using the whole dataset before splitting. This allows information about test data to leak into training.
        \textit{Spambase, LR}\\
        PF\_05 Augment early & Data augmentation is done before splitting data into train and test sets. This results in overlap between train and test sets. \textit{Wisconsin, SVM}\\
        PF\_06 Subject overlap & A dataset known to contain multiple samples per subject is split randomly. This results in overlap between train and test sets. \textit{PD, LR}\\
        PF\_07 Time series & The temporal relationship within time series data is not respected. This can result in meaningless models. \textit{Air traffic, LR}\\
        PF\_08 HPO no nest & When doing hyperparameter optimisation, standard CV is used rather than nested CV. This can lead to overfitting. \textit{Iris, RF}\\
        PF\_09 Ensemble leak & Overlap between data used to test ensemble and data used to train base models. Can result in misleading metrics. \textit{Iris, RF/LR/SVM}\\
        PF\_10 Embed leak & An embedding model is trained on the entire dataset. This allows information about test data to leak. \textit{Fashion MNIST, Autoencoder}\\
        PF\_11 Foundation leak & Foundation model applied to well-known benchmark. Likely to be in its original train data, leading to data contamination. \textit{IMDB, BERT}\\
        PF\_12 Bad features & Applying model expecting numerical features to categorical features. Invalidates assumptions, likely leading to a poor fit. \textit{AD, DT}\\
        PF\_13 Wrong task & Using regression model/metric to solve a classification task. This does not make sense and can lead to confusion. \textit{Titanic, LR}\\
        PF\_14 No HPO & Comparing models that have not been hyperparameter-tuned. Can result in misleading conclusions. \textit{California, RF/MLP/SVR}\\
        PF\_15 No regularize & Failing to use standard regularisation techniques for deep learning models. This can lead to overfitting. \textit{MNIST, CNN}\\
        PF\_16 Using accuracy & Using accuracy as the only metric on imbalanced data. Models that always predict the majority class can have high accuracy. \textit{Pima, LR}\\
        PF\_17 Ill-suited model & Complex model applied to data predictable using simple model. Leads to unnecessary complexity and resource usage. \textit{MNIST, LSTM}\\
        PF\_18 No baseline & Failing to compare against simple baselines. This can lead to invalid conclusions about the performance of a model. \textit{Air traffic, LSTM}\\
        PF\_19 Impute early & Doing mean imputation before splitting data. This allows information about test data to leak into training. \textit{Wisconsin, SVM}\\
        PF\_20 One metric & Failing to evaluate a model with multiple metrics. Can provide an incomplete/misleading picture of model performance. \textit{Wisconsin, RF}\\
        \bottomrule
    \end{tabularx}
    \caption{Catalogue of ML pitfalls considered in this work. Datasets and ML models shown in italics. Non-standard abbreviations --- California: California Housing Prices, Wisconsin: Breast Cancer Wisconsin, Air traffic: Airlines Traffic Passenger Statistics, Pima: Pima Indians Diabetes, PD: Parkinsons Telemonitoring, LR: logistic regression, AD: Adult dataset RF: random forest, DT: decision tree. Code samples for each pitfall can be found in the GitHub repository.}
    \label{tab:Mlpitfalls}
\end{table}

\subsection {Dataset of code samples containing pitfalls} \label{sec:methodology:samples}
Obtaining a large, diverse set of student programming assignments is a challenging task because of ethical issues related to student data privacy \citep{leinonen2024llmitationsincerestformdata}. Instead, we used a semisynthetic approach to generate code samples, in which a commercial-grade LLM (Google Gemini) is initially prompted to generate code, and a human expert then checks the functionality and correctness of the code. Beyond addressing the challenge of obtaining data, this approach is also motivated by several factors. First, it reflects the mode in which ML code is increasingly developed by students and practitioners, i.e.\ by querying an LLM and then refining the code as necessary. It also helps to avoid bias, since LLMs can be expected to replicate typical human coding behaviour, rather than be biased towards particular styles taught to specific cohorts. The use of synthetic code samples within computing education has been validated in \citep{leinonen2024llmitationsincerestformdata} where distributions of real and synthetic buggy code were found to be comparable.

In this study, LLMs are asked to generate Python code, since Python has become the \textit{de facto} standard for ML, both in education and in wider practice \citep{info11040193}. A single code sample was produced for each of the pitfalls listed in Table~\ref{tab:Mlpitfalls}. Whilst it would be possible to generate multiple samples per pitfall, this route was not taken due to its impact on resource usage and analytical overheads.

The LLM was not explicitly asked to generate incorrect code; rather it was prompted to generate code that solved a specific problem. This problem was chosen to be representative of the kind of problem a student would be given in a university-level ML course, based on an initial survey of such courses. Only solutions which exhibited the relevant pitfall were kept;  otherwise the LLM was re-prompted using a fresh session. In some cases, the error was introduced into the code. Note that the aim of this study is to measure the ability of LLMs to recognise pitfalls, not to generate them.  However, it is worth commenting on the fact that most samples generated by Gemini did exhibit the required pitfalls. Since Gemini was trained on code written by humans, the fact that it generated problematic code demonstrates how common these pitfalls must be in human-written samples. 

The GitHub repository\footnote{\url{https://github.com/ssk705/ML_Pitfalls}} provides access to the prompts, code samples, and results of the experiments. The source files in the repository are organized in two folders; the \texttt{src} folder provides the raw code, and commented versions can be found in the \texttt{src\_comments} folder. The commented versions were not used in the study, since our aim is to assess the ability of LLMs to assess code rather than accompanying comments, which may or may not be present in practice.

\begin{tcolorbox}[title=Repository Structure]
 \dirtree{%
.1 MLPitfalls.
.2 Results.
.2 dataset.
.2 src.
.2 src with comments.
}
\end{tcolorbox}

A three-part naming scheme is used for filenames in the \texttt{src} folder; for instance, \texttt{PF\_01\_LR\_CH.py} refers to the first pitfall (\texttt{PF\_01}) in Table~\ref{tab:Mlpitfalls}, the two-letter code for the algorithm (\texttt{LR}: Linear Regression), followed by a code for the dataset used (\texttt{CH}: California Housing Prices).

\subsection{LLMs used to identify pitfalls}  \label{sec:methodology:llms}

Using this dataset of code samples, we aimed to determine how well LLMs identify pitfalls in ML code, and the value of their feedback to student learners. We selected three open LLMs that were specifically trained to work with code. Their key characteristics are summarised in Table~\ref{tab:my_label}. At the time of writing, all ranked at the top of the Hugging Face Big Code Models leaderboard \citep{huggingfaceCodeModels}.
\begin{itemize}
    \item \textbf{CodeQwen1.5-7B} \citep{huggingfaceQwenCodeQwen157BHugging}, from Alibaba, is a long-context code LLM that supports multiple programming languages. 
    \item \textbf{CodeLlama-34b} \citep{codeLlama}, from Meta, is a larger (33.7 billion parameter) LLM designed for code synthesis and understanding.
    \item \textbf{DeepSeek-Coder-V2} \citep{deepseek} is intermediate in size to the other two, but is notable for using a mixture-of-experts model to improve performance.
\end{itemize}
As a baseline, we used OpenAI's closed LLM GPT-4, via GitHub Copilot \citep{githubGitHubCopilot}. Whilst this study particularly focuses on open LLMs, many students use commercial services, and the majority of these are based around GPT-4. This baseline allows us to evaluate the performance of open LLMs within this broader context. To reduce bias, we used a different closed LLM to the one (Gemini) used to generate code samples.

The open LLM environment was deployed using Ollama~\citep{ollama} on a Windows machine. GitHub Copilot interactions were carried out using Visual Studio Code.

\begin{table}[t]
     \centering
     \begin{tabular} 
    {@{}llrr@{}}\toprule
   
      Model & Model identifier & Parameters & Context length \\
         \midrule
         CodeQwen & CodeQwen1.5-7B  & 7.3 billion &  65536     \\        
         DeepSeek Coder & DeepSeek-Coder-V2 &15.7 billion  &163840 \\
         Code Llama & Codellama:34b-instruct & 33.7 billion &  16384   \\
           \bottomrule
     \end{tabular}
     \caption{Characteristics of the open LLMs}
     \label{tab:my_label}
 \end{table} 
 
\subsection{Prompt design}\label{sec:methodology:prompt}
Each of the LLMs from Section~\ref{sec:methodology:llms} were prompted with each of the code samples in turn. The following example shows how the prompts were structured:
\begin{tcolorbox}[title={Prompt for pitfall \#N}]
\texttt{Analyze the following machine learning code. The output should list the issues identified in the code with a detailed explanation and possible solutions.\\\#\#\#\#  \\ import pandas as pd \\
from sklearn$.$datasets import fetch\_california\_housing\\
from sklearn$.$linear\_model import LinearRegression\\
from sklearn$.$etrics import mean\_squared\_error, r2\_score\\
housing = fetch\_california\_housing(as\_frame=True)\\
df=pd$.$DataFrame(data=housing$.$data, columns=housing$.$feature\_names)\\
print(df.columns)\\
print(df)\\
df['MedHouseVal'] = housing.target\\
X = df.drop('MedHouseVal', axis=1)\\
y = df['MedHouseVal']\\
model = LinearRegression()\\
model.fit(X, y)\\
y\_pred = model.predict(X) \\
mse = mean\_squared\_error(y, y\_pred) \\
r2 = r2\_score(y, y\_pred) \\
print(f"Mean Squared Error: {mse:.2f}") \\
print(f"R-squared: {r2:.2f}") \\
\#\#\#\#}
\end{tcolorbox}

\subsection{Collection and evaluation of responses}\label{sec:methodology:evaluation}
In general, LLMs are non-deterministic, meaning that they can produce different responses each time they are given the same prompt. For this reason, each LLM is prompted five times for each pitfall example, and their responses are collected across these five repeats. In many LLMs, and especially open models which expose this setting to the user, the degree of non-determinism can be controlled via a temperature setting. This has a value between 0 and 1, with higher values producing greater diversity among responses to the same prompt. To account for this, we repeat prompting and collection of responses across five different temperature settings (0.3, 0.5, 0.7, 0.8, and 1.0). This leads to a set of 25 responses (5 repeats at each temperature) for each of the open models for each pitfall, and a total of 500 responses for each LLM across the whole pitfall dataset. For the closed model, GitHub Copilot, the temperature is not configurable, so only 5 responses are collected for each pitfall, totaling 100.

The correctness of LLM responses are measured using the recall metric:
\begin{equation} \text{Recall} = \text{True Positives} / (\text{True Positives} + \text{False Negatives}) \end{equation}
The recall rate across five repeats of the prompt is referred to as Recall@5. In this study, we are focusing on an LLM's ability to identify issues in code where these issues are present. This is characterised by True Positives, where a model correctly identifies the issue, and False Negatives, where they fail to identify the issue. We are not considering the situation where an LLM flags an issue that is not present (False Positives) or correctly identifies the lack of an issue (True Negatives). Whilst these are considerations within the use of LLMs, they are much less of a concern than the failure to spot an issue that is present, since unidentified issues pose a significant threat to the reliability and trustworthiness of machine learning systems.


For RQ2, we also consider whether the response provides knowledge of the mistake (KM) and knowledge of how to proceed (KH). For a response to be coded as KM, in addition to simply recognising the presence of a pitfall, the feedback should also adequately describe how the pitfall is present and what its consequences are, so that a learner can understand what they did wrong. For KH, it should also give an indication of what changes would need to be made to the code to remedy the pitfall. KM and KH are the types of elaborated feedback \citep{narciss2008feedback} most valuable to learners within this context. We do not consider whether KCR (knowledge of correct response) feedback is included --- whilst LLMs do sometimes generate corrected code, we considered this less valuable for student learners.

As an example, the following response for PF\_01 (no test) provides both KM and KH feedback, since it clearly articulates the nature of the problem and how it can be remedied: KM:  \textit{``The data is not split into training and testing sets before modeling. This means that the model may be overfitting to the training data, which can result in poor generalization performance on new, unseen data.''}
KH :  \textit{``To fix this, consider splitting the data into training and testing sets using scikit-learn's `train\_test\_split' function''}. 
 Such feedback will encourage the learner to reflect on the code, revisit relevant concepts and develop a deeper understanding of how theory and practice are connected.
 
The first author of this paper conducted the evaluation. 
    The LLM responses were validated to determine whether they identified the relevant pitfall (RQ1). The response was then checked for whether it included KM or KH feedback (RQ2), i.e. whether the underlying issue was adequately described, and whether guidance was provided on how to solve the issue.  For consistency, coding of responses was done by comparing against expert-written ground truth (provided in the repository). Since this is mostly an objective task, it does not require cross-coding. However, the second and third authors were involved in formulating the ground truth.
 
\section{Results and Discussion} \label{sec:results}
In this section, we present the results of our experiments. As noted above, each LLM was prompted with each of the code samples in turn, and the responses were checked for correctness and coded for their feedback. 
The 1600 raw responses are provided in the GitHub repository\footnote{\url{https://github.com/ssk705/ML_Pitfalls} \label{mlpitfalls}}. Below we focus on each of the two research questions in turn.

\subsection{RQ1: Reliability in identifying machine learning pitfalls}
We begin, in Section~\ref{sec:results:overall}, by considering the overall recall rates across the pitfalls and the comparative performance of each model. In Sections~\ref{sec:results:preprocessing}--\ref{sec:results:evaluation}, we then focus on groups of pitfalls occurring at the three phases of the ML pipeline: data preprocessing, model building, and evaluation.

\subsubsection{Overall performance and model comparison} \label{sec:results:overall}
\begin{table}[tb!]
\centering
\resizebox{\textwidth}{!}{%
\begin{tabular}{@{}lrrrrrr@{}}
\toprule
ML Pitfall & DeepSeek Coder & Code Llama & CodeQwen & Copilot & Mean recall & Rank \\
\midrule
PF\_16 Using accuracy & \cellcolor{green!90}92\% & \cellcolor{green!90}96\% & \cellcolor{green!70}72\% & \cellcolor{green!100}100\% & \cellcolor{green!90}90\% & 1 \\
PF\_20 One metric & \cellcolor{green!80}88\% & \cellcolor{green!90}96\% & \cellcolor{green!70}72\% & \cellcolor{green!100}100\% & \cellcolor{green!90}89\% & 2 \\
PF\_02 Reuse test & \cellcolor{green!80}84\% & \cellcolor{green!80}84\% & \cellcolor{green!80}88\% & \cellcolor{green!100}100\% & \cellcolor{green!90}89\% & 2 \\
PF\_01 No test & \cellcolor{green!80}88\% & \cellcolor{green!50}52\% & \cellcolor{green!80}80\% & \cellcolor{green!100}100\% & \cellcolor{green!80}80\% & 3 \\
PF\_09 Ensemble leak & \cellcolor{green!70}68\% & \cellcolor{green!40}40\% & \cellcolor{green!70}68\% & \cellcolor{green!100}100\% & \cellcolor{green!70}69\% & 4\\
PF\_14 No HPO & \cellcolor{green!70}72\% & \cellcolor{green!80}84\% & \cellcolor{green!80}80\% & \cellcolor{green!1}0\% & \cellcolor{green!50}59\% & 5 \\
PF\_13 Wrong task & \cellcolor{green!20}28\% & \cellcolor{green!40}44\% & \cellcolor{green!30}32\% & \cellcolor{green!100}100\% & \cellcolor{green!50}51\% & 6 \\
PF\_10 Embed leak & \cellcolor{green!40}44\% & \cellcolor{green!10}16\% & \cellcolor{green!20}28\% & \cellcolor{green!100}100\% & \cellcolor{green!50}47\% & 7 \\
PF\_07 Time series & \cellcolor{green!40}44\% & \cellcolor{green!20}24\% & \cellcolor{green!1}8\% & \cellcolor{green!100}100\% & \cellcolor{green!40}44\% & 8 \\
PF\_17 Ill-suited model & \cellcolor{green!1}4\% & \cellcolor{green!1}4\% & \cellcolor{green!10}16\% & \cellcolor{green!100}100\% & \cellcolor{green!30}31\% & 9 \\
PF\_03 Scale early & \cellcolor{green!10}12\% & \cellcolor{green!10}12\% & \cellcolor{green!1}0\% & \cellcolor{green!100}100\% & \cellcolor{green!30}31\% & 9 \\
PF\_12 Bad features & \cellcolor{green!60}56\% & \cellcolor{green!40}48\% & \cellcolor{green!20}20\% & \cellcolor{green!1}0\% & \cellcolor{green!30}31\% & 9 \\
PF\_15 No regularize & \cellcolor{green!20}20\% & \cellcolor{green!70}72\% & \cellcolor{green!10}16\% & \cellcolor{green!1}0\% & \cellcolor{green!20}27\% & 10 \\
PF\_08 HPO no nest & \cellcolor{green!1}8\% & \cellcolor{green!1}8\% & \cellcolor{green!1}8\% & \cellcolor{green!1}0\% & \cellcolor{green!1}6\% & 11 \\

PF\_05 Augment early & \cellcolor{green!1}8\% & \cellcolor{green!1}0\% & \cellcolor{green!1}4\% & \cellcolor{green!1}0\% & \cellcolor{green!1}3\% & 12 \\
PF\_19 Impute early & \cellcolor{green!10}12\% & \cellcolor{green!1}0\% & \cellcolor{green!1}0\% & \cellcolor{green!1}0\% & \cellcolor{green!1}3\% & 12 \\
PF\_04 Select early & \cellcolor{green!1}0\% & \cellcolor{green!10}12\% & \cellcolor{green!1}0\% & \cellcolor{green!1}0\% & \cellcolor{green!1}3\% & 12 \\
PF\_06 Subject overlap & \cellcolor{green!1}8\% & \cellcolor{green!1}0\% & \cellcolor{green!1}0\% & \cellcolor{green!1}0\% & \cellcolor{green!1}2\% & 13 \\
PF\_11 Foundation leak & \cellcolor{green!1}0\% & \cellcolor{green!1}0\% & \cellcolor{green!1}0\% & \cellcolor{green!1}0\% & \cellcolor{green!1}0\% & 14 \\
PF\_18 No baseline & \cellcolor{green!1}0\% & \cellcolor{green!1}0\% & \cellcolor{green!1}0\% & \cellcolor{green!1}0\% & \cellcolor{green!1}0\% & 14 \\
\midrule
Model mean recall & 37\% & 35\% & 30\% & 50\%  & 38\% \\
\bottomrule
\end{tabular}
}

\caption{Comparison of recall rates across pitfalls and LLMs, ordered by mean recall across LLMs.}
\label{table:summary}
\end{table}

As shown in Table~\ref{table:summary}, the mean recall rates of the models across all the pitfalls were 30\%, 37\%, 35\% and 50\% respectively for (in increasing order of model size) CodeQwen, DeepSeek Coder, Code Llama and Copilot. This implies that all of the LLMs fail to correctly identify pitfalls at least half of the time. It also shows a generally increasing trend with model size, with the notable exception of the 15.7 billion parameter DeepSeek Coder slightly outperforming the larger 33.7 billion parameter Code Llama. The closed Copilot LLM does, on this metric at least, beat the open models. Nevertheless, the performance of open and closed models is closer than might be expected given that the LLM used by Copilot is an order of magnitude larger than these open models. However, the recall rates are quite different in that Copilot has recall rates of either 100\% or 0\%, as opposed to the more variable responses of the open models. This is presumably due to a low temperature setting in the Copilot service, favouring consistency over diversity of responses. However, it does mean that Copilot misses 50\% of the pitfalls completely, whereas DeepSeek Coder only misses 15\% completely. We further discuss the effect of temperature in Section~\ref{sec:results:temperature}.

 
Table~\ref{table:summary} also ranks the pitfalls in order of mean recall rates achieved across the four LLMs. The first five of these (ranks 1--4) have relatively consistent recall rates across the models, and include widely known pitfalls such as failing to use the test set correctly and using accuracy with imbalanced data. It would not be surprising to see these pitfalls discussed in an introductory machine learning course, suggesting that LLMs could be used to support learning at this stage. The majority of the entries in the middle group (ranks 11 onward) are concerned with inappropriate model or component choices. This is generally a concern in the early stages of systems thinking, when students have developed theoretical knowledge of ML models and are beginning to apply them to specific data sets. The patchy ability of LLMs to spot these pitfalls is hence a concern. The last seven in the table (ranks 11 onward) all have consistently low recall rates across the four models. Almost all of these would lead to information leaks, meaning they are an important source of error once students are starting to apply ML to real world problems, and therefore something we would like machine learning students to have a good comprehension of. From this perspective, it is concerning that they have largely been missed by the LLMs, questioning the ability of current LLMs to support learning during later education and the transition to practice.

Another notable observation is the deviation in recall for all models between different pipeline stages. At the preprocessing stage (PF\_03--PF\_07 and PF\_19), Copilot and the best performing open model (DeepSeek Coder) achieve mean recall rates of 33\% and 14\%. At the model building stage (PF\_08--15 and PF\_17), this grows to 44\% and 33\%, respectively, and at the evaluation stage (PF\_01-02, PF\_16, PF\_18 and PF\_20), 70\% and 80\%. This suggests a very uneven ability to spot pitfalls across the ML pipeline, with very low rates at the preprocessing phase, where mistakes can have a particularly large impact on an ML project. We dig deeper into the ability of models to recognise pitfalls at these three stages in the following sections.



\subsubsection{Pitfalls associated with data preprocessing} \label{sec:results:preprocessing}

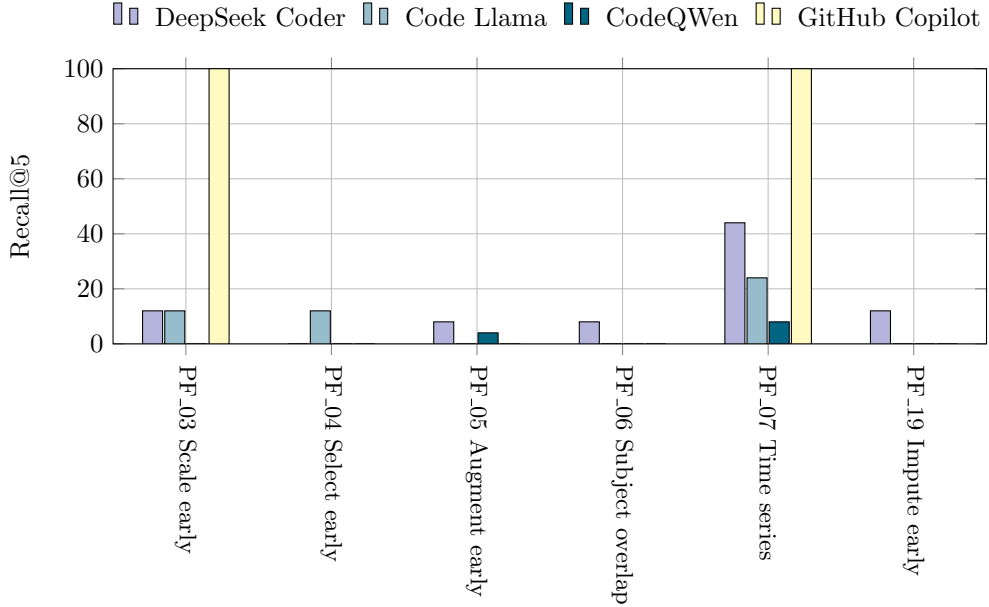
\begin{figure} [h]
\centering
\begin{tikzpicture}
\begin{axis}[
    legend style={
        at={(0.5,1.1)}, 
        anchor=south, 
        legend columns=-1,
        column sep=1ex,
        draw=none
    },
    ybar=2*\pgflinewidth,
    symbolic x coords={ PF\_03 Scale early, PF\_04 Select early, PF\_05 Augment early, PF\_06 Subject overlap, PF\_07 Time series, PF\_19 Impute early},
    bar width= 7.5pt,
    xtick=data,
    ylabel={Recall@5},
    ymin=0, ymax=100,
    grid=major,
    width=\textwidth,
    height=0.4\textwidth,
    xticklabel style={rotate=270, anchor=west, font=\small},
]



\addplot[fill=lightblue] coordinates {(PF\_03 Scale early, 12) (PF\_04 Select early, 0) (PF\_05 Augment early, 8)
                      (PF\_06 Subject overlap, 8) (PF\_07 Time series, 44) (PF\_19 Impute early, 12) };
 \addplot[fill=pastel] coordinates {(PF\_03 Scale early, 12) (PF\_04 Select early, 12) (PF\_05 Augment early, 0)
                      (PF\_06 Subject overlap, 0) (PF\_07 Time series, 24) (PF\_19 Impute early, 0) };

\addplot [fill=darkerblue] coordinates { (PF\_03 Scale early, 0) (PF\_04 Select early, 0) (PF\_05 Augment early, 4) (PF\_06 Subject overlap, 0) (PF\_07 Time series, 8) (PF\_19 Impute early, 0) };

\addplot [fill=yellow!30] coordinates { (PF\_03 Scale early, 100) (PF\_04 Select early, 0) (PF\_05 Augment early, 0) (PF\_06 Subject overlap, 0) (PF\_07 Time series, 100) (PF\_19 Impute early, 0) };

\legend{DeepSeek Coder,Code Llama,CodeQWen,GitHub Copilot}                      
\end{axis}
\end{tikzpicture}
\caption{Summary of accuracy@5 scores across different temperatures for DeepSeek Coder, Code Llama, CodeQWen and GitHub Copilot for pitfalls related to the data preprocessing stage of a machine learning pipeline (PF\_03 to PF\_07 and PF\_19) }
\label{pitfall1to7and19}
\end{figure}
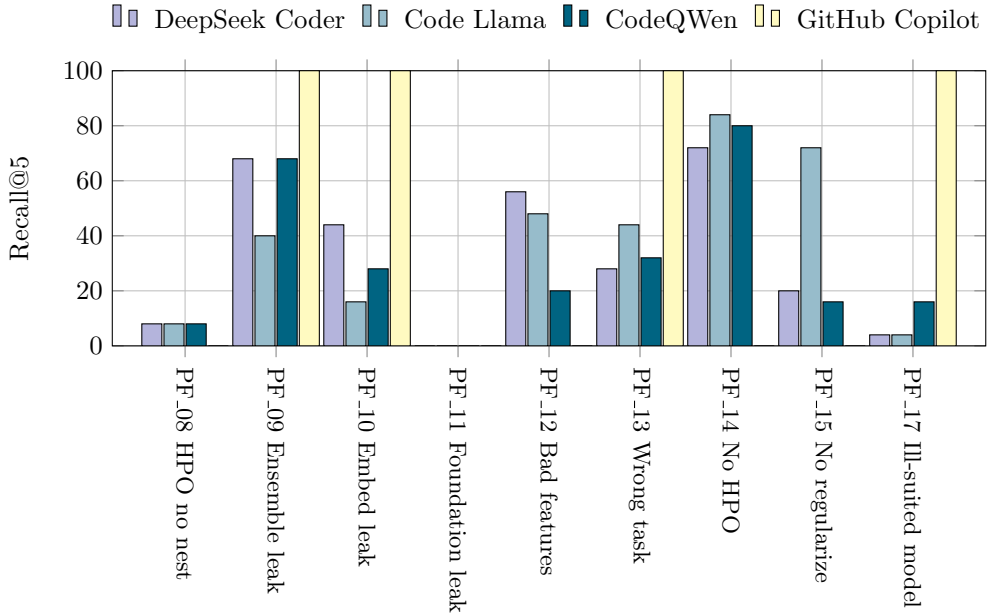
\begin{figure} [h]
\centering
\begin{tikzpicture}
\begin{axis}[
    legend style={
        at={(0.5,1.1)}, 
        anchor=south, 
        legend columns=-1,
        column sep=1ex,
        draw=none
    }, 
    ybar=2*\pgflinewidth,
 symbolic x coords={PF\_08 HPO no nest, PF\_09 Ensemble leak, PF\_10 Embed leak, PF\_11 Foundation leak, PF\_12 Bad features, PF\_13 Wrong task, PF\_14 No HPO, PF\_15 No regularize, PF\_17 Ill-suited model},    
    bar width= 7.5pt,
    xtick=data,
    ylabel={Recall@5},
    ymin=0, ymax=100,
    grid=major,
    width=\textwidth,
    height=0.4\textwidth,
    xticklabel style={rotate=270,anchor=west, font=\small},
]
\addplot[fill = lightblue] coordinates {(PF\_08 HPO no nest, 8) (PF\_09 Ensemble leak, 68) (PF\_10 Embed leak, 44) (PF\_11 Foundation leak, 0) (PF\_12 Bad features, 56) (PF\_13 Wrong task, 28) (PF\_14 No HPO, 72) (PF\_15 No regularize, 20) (PF\_17 Ill-suited model, 4)};
 \addplot [fill =pastel] coordinates {(PF\_08 HPO no nest, 8) (PF\_09 Ensemble leak, 40) (PF\_10 Embed leak, 16) (PF\_11 Foundation leak, 0) (PF\_12 Bad features, 48)
                      (PF\_13 Wrong task, 44) (PF\_14 No HPO, 84) (PF\_15 No regularize, 72) (PF\_17 Ill-suited model, 4)};
\addplot [fill = darkerblue] coordinates {(PF\_08 HPO no nest, 8) (PF\_09 Ensemble leak, 68) (PF\_10 Embed leak, 28) (PF\_11 Foundation leak, 0) (PF\_12 Bad features, 20) (PF\_13 Wrong task, 32) (PF\_14 No HPO, 80)  (PF\_15 No regularize, 16)  (PF\_17 Ill-suited model, 16) };

\addplot [fill=yellow!30] coordinates {(PF\_08 HPO no nest, 0) (PF\_09 Ensemble leak, 100) (PF\_10 Embed leak, 100) (PF\_11 Foundation leak, 0) (PF\_12 Bad features, 0) (PF\_13 Wrong task, 100) (PF\_14 No HPO, 0)  (PF\_15 No regularize, 0) (PF\_17 Ill-suited model, 100)};

\legend{DeepSeek Coder,Code Llama,CodeQWen,GitHub Copilot}                      
\end{axis}
\end{tikzpicture}
\caption{Summary of recall@5 scores across different temperatures for DeepSeek Coder, Code Llama, CodeQwen, and GitHub Copilot  for pitfalls related to the model building stage of a machine learning pipeline (PF\_08 to PF\_15 and PF\_17) }
\label{pitfall8to15}
\end{figure}

\begin{figure} [h]
\centering
\begin{tikzpicture}
\begin{axis}[
     legend style={
        at={(0.5,1.1)}, 
        anchor=south, 
        legend columns=-1,
        column sep=1ex,
        draw=none
    },
    ybar=2*\pgflinewidth,
    symbolic x coords={PF\_01 No test, PF\_02 Reuse test, PF\_16 Using accuracy, PF\_18 No baseline, PF\_20 One metric},
    bar width= 7.5pt,
    xtick=data,
    ylabel={Recall@5},
    ymin=0, ymax=100,
    grid=major,
    width=\textwidth,
    height=0.4\textwidth,
    xticklabel style={rotate=270, anchor=west, font=\small},
]
\addplot [fill = lightblue] coordinates {(PF\_01 No test, 88) (PF\_02 Reuse test, 84)(PF\_16 Using accuracy, 92) 
                      (PF\_18 No baseline, 0)   (PF\_20 One metric, 88) };
 \addplot [fill =pastel] coordinates {(PF\_01 No test, 52) (PF\_02 Reuse test, 84) (PF\_16 Using accuracy, 96)  (PF\_18 No baseline, 0)   (PF\_20 One metric, 96)};

\addplot [fill = darkerblue] coordinates {(PF\_01 No test, 80) (PF\_02 Reuse test, 88) (PF\_16 Using accuracy, 72) (PF\_18 No baseline, 0)   (PF\_20 One metric, 72)  };

\addplot [fill=yellow!30] coordinates {(PF\_01 No test, 100) (PF\_02 Reuse test, 100) (PF\_16 Using accuracy, 100)  (PF\_18 No baseline, 0)   (PF\_20 One metric, 100)  };

\legend{DeepSeek Coder,Code Llama,CodeQWen,GitHub Copilot}                      
\end{axis}
\end{tikzpicture}
\caption{Summary of recall@5 scores across different temperatures for DeepSeek Coder, Code Llama, CodeQwen, and GitHub Copilot for pitfalls related to the model evaluation stage of a machine learning pipeline (PF\_01, PF\_02, PF\_16, PF\_18 and PF\_20)}
\label{pitfall16171820}
\end{figure}
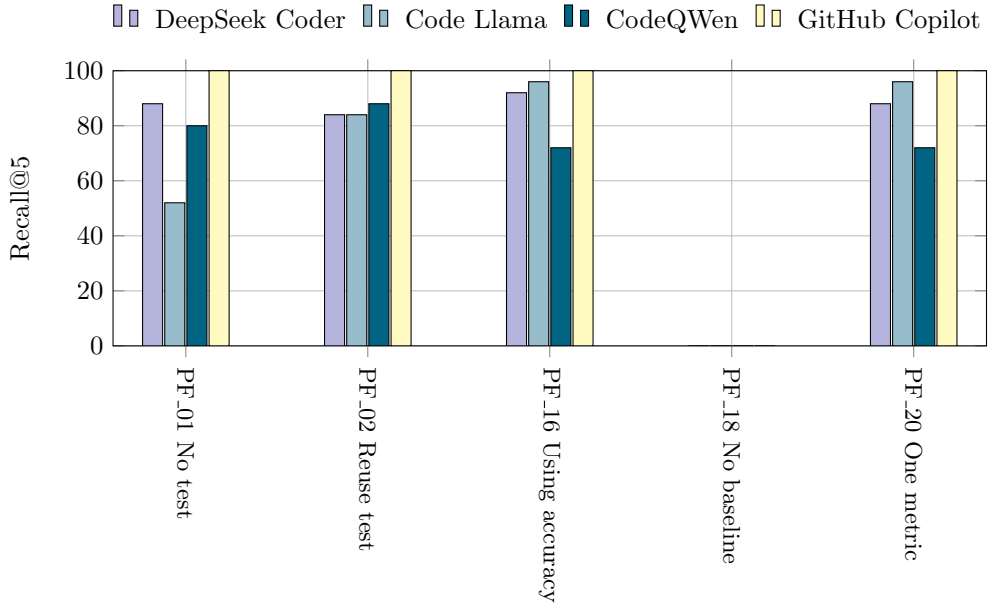

Figure~\ref{pitfall1to7and19} shows the recall rates for pitfalls at this stage of the ML pipeline.
PF\_03 to PF\_05,  and PF\_19 are all concerned with the application of data-dependent operations too early in the ML pipeline, potentially resulting in information leakage. These errors have been commonly observed in published scientific ML pipelines \citep{KAPOOR2023100804}, so it is notable that all the models struggle to pick up on them. The only exception is for PF\_03, which is correctly identified by the closed Copilot model. This is the simplest of the four, and is concerned with scaling; a sample correct response was \textit{``Split the data first, then apply scaling to the training and testing sets separately''.} A typical misleading response from the other models was that scaling is unnecessary as the Iris dataset features are already normally distributed; the latter may be true, but it does not relate to whether or not scaling is required.

The open models had low recall rates across all four of these pitfalls, with DeepSeek Coder doing slightly better than the others. However, Code Llama was the only model to pick up on PF\_04 at all (with only 12\% recall). This is concerned with doing feature selection before partitioning, which is a common error. An example of a potentially misleading response for this pitfall was to perform feature selection before scaling, something that is debatable and potentially suboptimal. An example of an uninformative response, this time for PF\_19, was \textit{``Only the training set is scaled (`scaler.fit\_transform(X\_train)'), while the test set is only transformed (`scaler.transform(X\_test)')''}, which shows a lack of understanding of what these function calls are for.

The models also demonstrated poor performance on PF\_06, which is concerned with subject overlap between train and test sets, again leading to information leakage. DeepSeek Coder was the only model to pick up on this, with a recall of 8\%. This is admittedly a difficult pitfall to spot in code; however, the dataset name should give some indication to the model, since both this dataset specifically and medical datasets more generally exhibit the potential issue of containing multiple samples per subject. The reference to ``patient\_id'' in the code also gives a clue, though this seemed to act more as a distractor, with common responses such as \textit{``The column 'patient\_id\#' should be removed from the feature set as it does not seem to contribute any meaningful information for prediction based on the given data description''}. This illustrates the commonly observed property that LLMs are sensitive to the exact wording of their input.


For PF\_07, concerned with partitioning time series data, two of the models were able to recognise this with reasonable consistency: Copilot and DeepSeek Coder, the latter with a recall of 44\%. The others had very low recall, which is perhaps surprising given that randomly shuffling time series data is a fundamental error that is not hard to spot in code.



\subsubsection{Pitfalls associated with model building}
 

Figure~\ref{pitfall8to15} shows the recall rates for pitfalls at this stage of the ML pipeline. PF\_08 and PF\_14 are both concerned with hyperparameter optimisation (HPO). \citet{zimmermann2023common} noted the lack of HPO as a common issue in student ML code. From this perspective, it is reassuring to see that the open models generally do well at recognising the absence of HPO (FP\_14), though worrying that Copilot does not pick up on this. However, all models did poorly at PF\_08, which is concerned with the incorrect implementation of HPO. Less than 10\% of responses from the closed models noted that the lack of nesting in cross-validation could lead to overfitting, and Copilot did not recognise this issue at all.


PF\_09--11 are all concerned with information leaks when using multiple models. Performance on PF\_09, which focused on leaks during the training of ensemble models, was generally good, with only Code Llama dropping below 50\% recall. For PF\_10, concerned with leaks due to pretraining of embedding models, DeepSeek Coder and Copilot did relatively well, but the other open models had low recall rates. PF\_11, regarding data contamination when using a pretrained foundation model, was not identified by any of the models. This is concerning, since use of foundation models has become standard practice in ML, and the scope for contamination in the example should be apparent from the use of the widely-used BERT model in combination with a widely-used NLP dataset.


PF\_12 and PF\_13 are concerned with inappropriate model choices that are likely to lead to a poor fit. In PF\_12, a decision tree is applied to non-continuous categorical features that are being incorrectly treated as numerical. Only DeepSeek Coder is able to spot it more than half the time. Copilot has zero recall. PF\_13 involves a regression model being used to solve a classification task. This time, Copilot recognised the issue, and Code Llama did best out of the open models, though with only 44\% recall.

PF\_15 and PF\_17 focus on the use of unnecessarily complex neural network models, which may lead to overfitting and excess resource usage. In the case of PF\_17, an LSTM (a model for processing sequential data) is being inappropriately applied to the simple non-sequential Iris dataset. Copilot correctly picks up on this, but it was largely missed by the open models. PF\_15 concerns the lack of regularisation, the use of which is standard practice when training neural networks in order to discourage overfitting. Only Code Llama spots this one relatively consistently, with 72\% recall. Copilot does not spot it at all.


\subsubsection{Pitfalls associated with model evaluation} \label{sec:results:evaluation}

Figure~\ref{pitfall16171820} shows the recall rates for pitfalls at this stage of the ML pipeline.
PF\_01 and 02 are both concerned with errors of practice around test sets: not using one in the former, and using the same one multiple times in the latter. These are fundamental errors, so it is good to see that the recall rates are high across all models, with the slight exception of Code Llama on PF\_01. In addition to spotting the target pitfall, the models also suggested other improvements related to test data usage. For example, in one instance,  DeepSeek Coder responded with \textit{``The model is trained and evaluated multiple times without any validation or cross-validation, which can lead to overly optimistic performance estimates"}.

PF\_16 and PF\_20 are both to do with choice of metrics; the inappropriate use of accuracy with imbalanced datasets for the former, and the failure to use multiple metrics for the latter. All the models do well on both of these. PF\_16, in particular, is recognised with over 90\% recall by three of the models, with only CodeQwen at 72\%.

PF\_18, on the other hand, has zero recall across all the models. This concerns the lack of baseline comparisons, the absence of which can lead to inappropriate conclusions about achievable performance and the suitability of model choices. This is a particular issue within time series prediction (around which this particular code sample is framed) where it has led to the use of needlessly complex models \citep{hewamalage2023forecast}.



\subsection{Effect of temperature on model performance} \label{sec:results:temperature}

Temperature is a setting that affects the non-determinism of LLMs. Low value temperatures lead to more consistent outputs across invocations of the model, and high temperatures lead to more variety. In the results above, we have measured performance across a range of temperature settings and then averaged these. In this section, we take a closer look at the effect of temperature.

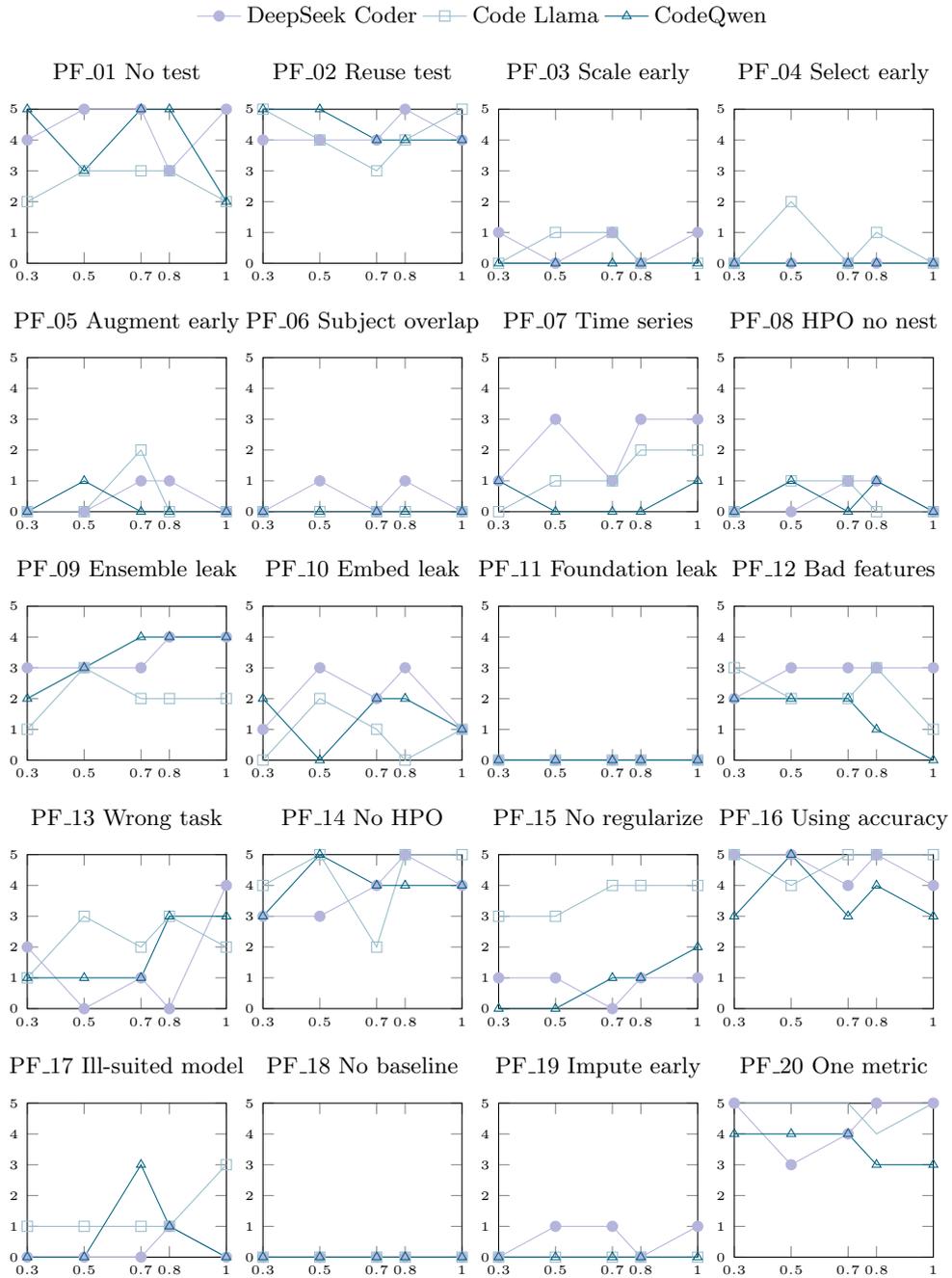
\begin{figure}
\centering
\begin{tikzpicture}
\begin{groupplot}[
    group style={
        group size= 4 by 5, 
        horizontal sep=0.5cm, 
        vertical sep=1.3cm, 
    },
    tick label style={font=\tiny},
    width=4.3cm, 
    height=3.7cm, 
    xlabel={},
    ylabel={ },    
    xmin=0.3, xmax=1,  
    ymin=0, ymax=5,  
    xtick={0.3,0.5,0.7,0.8,1},  
    ytick={0,1,2,3,4,5},
    every axis label/.append style={font=\tiny},
    title style={font=\small, text depth=0.5ex},
    legend style={
        font=\small,     
        at={($(0,0)+(2.2cm,3.4cm)$)},  
        anchor=west,  
        legend columns=3,
        align=center,
        draw=none
    }
]
\nextgroupplot[title={PF\_01 No test}]
\addplot[color= {lightblue}, mark=*] coordinates {(0.3,4) (0.5,5) (0.7,5) (0.8,3) (1,5)};
\addlegendentry{DeepSeek Coder}
\addplot[color= {pastel}, mark=square] coordinates {(0.3,2) (0.5,3) (0.7,3) (0.8,3) (1,2)};
\addlegendentry{Code Llama}
\addplot[color ={darkerblue}, mark=triangle] coordinates {(0.3,5) (0.5,3) (0.7,5) (0.8,5) (1,2)};
\addlegendentry{CodeQwen}

\nextgroupplot[title={PF\_02 Reuse test}]
\addplot[color= {lightblue}, mark=*] coordinates {(0.3,4) (0.5,4) (0.7,4) (0.8,5) (1,4)};
 
\addplot[color= {pastel}, mark=square] coordinates {(0.3,5) (0.5,4) (0.7,3) (0.8,4) (1,5)};
 
\addplot[color ={darkerblue}, mark=triangle] coordinates {(0.3,5) (0.5,5) (0.7,4) (0.8,4) (1,4)};

\nextgroupplot[title={PF\_03 Scale early}]
\addplot[color= {lightblue}, mark=*] coordinates {(0.3,1) (0.5,0) (0.7,1) (0.8,0) (1,1)};
 
\addplot[color= {pastel}, mark=square] coordinates {(0.3,0) (0.5,1) (0.7,1) (0.8,0) (1,0)};
 
\addplot[color ={darkerblue}, mark=triangle] coordinates {(0.3,0) (0.5,0) (0.7,0) (0.8,0) (1,0)};

\nextgroupplot[title={PF\_04 Select early}]
\addplot[color= {lightblue}, mark=*] coordinates {(0.3,0) (0.5,0) (0.7,0) (0.8,0) (1,0)};
 
\addplot[color= {pastel}, mark=square] coordinates {(0.3,0) (0.5,2) (0.7,0) (0.8,1) (1,0)};
 
\addplot[color ={darkerblue}, mark=triangle] coordinates {(0.3,0) (0.5,0) (0.7,0) (0.8,0) (1,0)};
\nextgroupplot[title={PF\_05 Augment early}]

\addplot[color= {lightblue}, mark=*] coordinates {(0.3,0) (0.5,0) (0.7,1) (0.8,1) (1,0)};
 
\addplot[color= {pastel}, mark=square] coordinates {(0.3,0) (0.5,0) (0.7,2) (0.8,0) (1,0)};
 
\addplot[color ={darkerblue}, mark=triangle] coordinates {(0.3,0) (0.5,1) (0.7,0) (0.8,0) (1,0)};
\nextgroupplot[title={PF\_06 Subject overlap}]
\addplot[color= {lightblue}, mark=*] coordinates {(0.3,0) (0.5,1) (0.7,0) (0.8,1) (1,0)};
 
\addplot[color= {pastel}, mark=square] coordinates {(0.3,0) (0.5,0) (0.7,0) (0.8,0) (1,0)};
 
\addplot[color ={darkerblue}, mark=triangle] coordinates {(0.3,0) (0.5,0) (0.7,0) (0.8,0) (1,0)};

\nextgroupplot[title={PF\_07 Time series}]
\addplot[color= {lightblue}, mark=*] coordinates {(0.3,1
) (0.5,3) (0.7,1) (0.8,3) (1,3)};
 
\addplot[color= {pastel}, mark=square] coordinates {(0.3,0) (0.5,1) (0.7,1) (0.8,2) (1,2)};
 
\addplot[color ={darkerblue}, mark=triangle] coordinates {(0.3,1) (0.5,0) (0.7,0) (0.8,0) (1,1)};

\nextgroupplot[title={PF\_08 HPO no nest }]
\addplot[color= {lightblue}, mark=*] coordinates {(0.3,0) (0.5,0) (0.7,1) (0.8,1) (1,0)};
 
\addplot[color= {pastel}, mark=square] coordinates {(0.3,0) (0.5,1) (0.7,1) (0.8,0) (1,0)};
 
\addplot[color ={darkerblue}, mark=triangle] coordinates {(0.3,0) (0.5,1) (0.7,0) (0.8,1) (1,0)};
\nextgroupplot[title={PF\_09  Ensemble leak}]
\addplot[color= {lightblue}, mark=*] coordinates {(0.3,3) (0.5,3) (0.7,3) (0.8,4) (1,4)};
 
\addplot[color= {pastel}, mark=square] coordinates {(0.3,1) (0.5,3) (0.7,2) (0.8,2) (1,2)};
 
\addplot[color ={darkerblue}, mark=triangle] coordinates {(0.3,2) (0.5,3) (0.7,4) (0.8,4) (1,4)};
\nextgroupplot[title={PF\_10  Embed leak}]
\addplot[color= {lightblue}, mark=*] coordinates {(0.3,1) (0.5,3) (0.7,2) (0.8,3) (1,1)};
 
\addplot[color= {pastel}, mark=square] coordinates {(0.3,0) (0.5,2) (0.7,1) (0.8,0) (1,1)};
 
\addplot[color ={darkerblue}, mark=triangle] coordinates {(0.3,2) (0.5,0) (0.7,2) (0.8,2) (1,1)};
\nextgroupplot[title={PF\_11 Foundation leak}]
\addplot[color= {lightblue}, mark=*] coordinates {(0.3,0) (0.5,0) (0.7,0) (0.8,0) (1,0)};
 
\addplot[color= {pastel}, mark=square] coordinates {(0.3,0) (0.5,0) (0.7,0) (0.8,0) (1,0)};
 
\addplot[color ={darkerblue}, mark=triangle] coordinates {(0.3,0) (0.5,0) (0.7,0) (0.8,0) (1,0)};

\nextgroupplot[title={PF\_12  Bad features}]
\addplot[color= {lightblue}, mark=*] coordinates {(0.3,2) (0.5,3) (0.7,3) (0.8,3) (1,3)};
 
\addplot[color= {pastel}, mark=square] coordinates {(0.3,3) (0.5,2) (0.7,2) (0.8,3) (1,1)};
 
\addplot[color ={darkerblue}, mark=triangle] coordinates {(0.3,2) (0.5,2) (0.7,2) (0.8,1) (1,0)};
\nextgroupplot[title={PF\_13 Wrong task}]
\addplot[color= {lightblue}, mark=*] coordinates {(0.3,2) (0.5,0) (0.7,1) (0.8,0) (1,4)};
 
\addplot[color= {pastel}, mark=square] coordinates {(0.3,1) (0.5,3) (0.7,2) (0.8,3) (1,2)};
 
\addplot[color ={darkerblue}, mark=triangle] coordinates {(0.3,1) (0.5,1) (0.7,1) (0.8,3) (1,3)};
\nextgroupplot[title={PF\_14 No HPO}]
\addplot[color= {lightblue}, mark=*] coordinates {(0.3,3) (0.5,3) (0.7,4) (0.8,5) (1,4)};
 
\addplot[color= {pastel}, mark=square] coordinates {(0.3,4) (0.5,5) (0.7,2) (0.8,5) (1,5)};
 
\addplot[color ={darkerblue}, mark=triangle] coordinates {(0.3,3) (0.5,5) (0.7,4) (0.8,4) (1,4)};
\nextgroupplot[title={PF\_15 No regularize}]
\addplot[color= {lightblue}, mark=*] coordinates {(0.3,1) (0.5,1) (0.7,0) (0.8,1) (1,1)};
 
\addplot[color= {pastel}, mark=square] coordinates {(0.3,3) (0.5,3) (0.7,4) (0.8,4) (1,4)};
 
\addplot[color ={darkerblue}, mark=triangle] coordinates {(0.3,0) (0.5,0) (0.7,1) (0.8,1) (1,2)};
\nextgroupplot[title={PF\_16  Using accuracy}]
\addplot[color= {lightblue}, mark=*] coordinates {(0.3,5) (0.5,5) (0.7,4) (0.8,5) (1,4)};
 
\addplot[color= {pastel}, mark=square] coordinates {(0.3,5) (0.5,4) (0.7,5) (0.8,5) (1,5)};
 
\addplot[color ={darkerblue}, mark=triangle] coordinates {(0.3,3) (0.5,5) (0.7,3) (0.8,4) (1,3)};
\nextgroupplot[title={PF\_17 Ill-suited model}]
\addplot[color= {lightblue}, mark=*] coordinates {(0.3,0) (0.5,0) (0.7,0) (0.8,1) (1,0)};
 
\addplot[color= {pastel}, mark=square] coordinates {(0.3,1) (0.5,1) (0.7,1) (0.8,1) (1,3)};
 
\addplot[color ={darkerblue}, mark=triangle] coordinates {(0.3,0) (0.5,0) (0.7,3) (0.8,1) (1,0)};

 \nextgroupplot[title={PF\_18 No baseline}]
\addplot[color= {lightblue}, mark=*] coordinates {(0.3,0) (0.5,0) (0.7,0) (0.8,0) (1,0)};
\addplot[color= {pastel}, mark=square] coordinates {(0.3,0) (0.5,0) (0.7,0) (0.8,0) (1,0)};
\addplot[color ={darkerblue}, mark=triangle] coordinates {(0.3,0) (0.5,0) (0.7,0) (0.8,0) (1,0)};
\nextgroupplot[title={PF\_19 Impute early}]
\addplot[color= {lightblue}, mark=*] coordinates {(0.3,0) (0.5,1) (0.7,1) (0.8,0) (1,1)};
 
\addplot[color= {pastel}, mark=square] coordinates {(0.3,0) (0.5,0) (0.7,0) (0.8,0) (1,0)};
 
\addplot[color ={darkerblue}, mark=triangle] coordinates {(0.3,0) (0.5,0) (0.7,0) (0.8,0) (1,0)};
\nextgroupplot[title={PF\_20 One metric}]
\addplot[color= {lightblue}, mark=*] coordinates {(0.3,5) (0.5,3) (0.7,4) (0.8,5) (1,5)};
\addplot[color= {pastel}] coordinates {(0.3,5) (0.5,5) (0.7,5) (0.8,4) (1,5)};
\addplot[color ={darkerblue}, mark=triangle] coordinates {(0.3,4) (0.5,4) (0.7,4) (0.8,3) (1,3)};

\end{groupplot}
\end{tikzpicture}
\caption{Performance of the open LLMs at temperature values 0.3, 0.5, 0.7, 0.8 and 1.0. The x-axis shows the temperature, and the y-axis shows the number of correct responses across 5 invocations.}

\label{TempAcc_modelevaluation}
\end{figure}

Figure~\ref{TempAcc_modelevaluation} shows the effect of temperature on the ability of the open LLMs to correctly identify pitfalls. On the whole, this suggests that there is no value of temperature that consistently works well across all models and pitfalls, and this justifies our approach of averaging recall rates across temperatures. However, if we restrict the analysis to those ranked in the top six in Table~\ref{table:summary}, where a greater proportion of interactions were successful, there is some tendency for larger values of temperature to be more optimal. For example, considering the mean optimal temperatures across models, for PF\_01, PF\_09, PF\_14, PF\_16 and PF\_20, there are optima at temperature settings of 0.8 and 1.0. Nevertheless, there are also cases where low temperature settings are more optimal. For some pitfalls, the temperature has a relatively small effect (e.g.\ PF\_09 and PF\_20). For others, the effect is larger. However, the choice of model is also a significant factor here. For instance, for PF\_16, two models remain optimal across temperature settings, but the third (CodeQwen) varies considerably.

If we take a look at pitfalls at the other end of the recall spectrum, it is again hard to spot consistent patterns. However, one notable observation is that for most of these, the highest temperature setting (1.0) is associated with lower recall. This may suggest that, given the models are already struggling with these pitfalls, then adding variability to their response may take them further away from the truth.

\subsection{RQ2: How useful is the feedback generated?}\label{RQ2}

Each response from the four LLMs was subsequently coded as KM (knowledge of mistake) or KH (knowledge of how to proceed) according to the procedure outlined in Section~\ref{sec:methodology:evaluation}. Table~\ref{tab:kmkh} shows the results, indicating which proportion of responses included these two types of feedback. Significantly, the figures for KM are identical to those reported in the previous section, indicating that on every occasion where an LLM identified a pitfall, it also provided information about why the code was erroneous. Compared to earlier approaches, this highlights one of the benefits of LLMs, in that some kind of feedback almost always comes included as default in their response.

The final column of Table~\ref{tab:kmkh} for each model shows which proportion of responses with KM also include KH, or equally which proportion of correctly identified pitfalls also included information about how the code should be changed to remove the error. These figures are generally very high, indicating that most correct responses also included information that would guide the learner towards a correct solution. Again, it appears to be a fairly default behaviour of LLMs to provide relevant feedback, and this also includes guidance to learners on how to proceed. For Copilot in particular, this kind of feedback is provided in all cases where the pitfall was correctly recognised.

For the open models, there are several pitfalls for which KH feedback is not routinely provided. In most cases, this corresponds to low recall rates, where the LLM also struggles to identify the pitfall; in this case, note that the figures in the final column are also less reliable due to small sample sizes. However, a general observation with these cases is that the LLM often provides a list of potential issues in the code, without providing much detail about any of them, perhaps reflecting its low confidence. PF\_09 notably has low values in the final column despite reasonable recall rates; in this case, the models do report an issue with different test data splits being used for different base models, but they do not adequately relate this to feedback on how to avoid data leaks in this situation. This highlights the existence of cases where an LLM can solve the easier task of recognising an issue, but not the harder task of resolving it.

A further observation is that feedback varies considerably in terms of length and detail. Some responses only indicate the general method for remedying an issue, whereas others include specific information about function and library usage, and others provide full replacement code. In practice, the form of feedback does influence its pedagogical role. Providing replacement code, in particular, has the potential to impede the learning process by encouraging learners to copy-and-paste solutions rather than developing comprehension of the underlying issue. Whilst it would be possible to further analyse the responses based on the exact nature of the feedback, in practise this may be ineffectual. This is because LLMs are sensitive to the formulation of their prompts; for instance, adding a further instruction like ``Provide information about how to fix the error without providing code'' could likely change the nature of the response, making it challenging to decouple the influence of prompt wording from that of the underlying model.

\begin{table}[tb!]
\centering
\resizebox{\textwidth}{!}{%
\begin{tabular}{@{}lrrr|rrr|rrr|rrr}
\toprule
ML Pitfall & \multicolumn{3}{c}{DeepSeek Coder} & \multicolumn{3}{c}{Code Llama} & \multicolumn{3}{c}{CodeQwen} & \multicolumn{3}{c}{Copilot} \\
&
\multicolumn{1}{c}{KM} & \multicolumn{1}{c}{KH} & \multicolumn{1}{c}{\sfrac{KH}{KM}} &
\multicolumn{1}{c}{KM} & \multicolumn{1}{c}{KH} & \multicolumn{1}{c}{\sfrac{KH}{KM}} &
\multicolumn{1}{c}{KM} & \multicolumn{1}{c}{KH} & \multicolumn{1}{c}{\sfrac{KH}{KM}} &
\multicolumn{1}{c}{KM} & \multicolumn{1}{c}{KH} & \multicolumn{1}{c}{\sfrac{KH}{KM}}
\\
\midrule
PF\_01 No test &
\cellcolor{green!80}  88\% & \cellcolor{green!80}  88\% & \cellcolor{green!100}100\% &
\cellcolor{green!50}  52\% & \cellcolor{green!30}  32\% & \cellcolor{green!60}  62\% &
\cellcolor{green!80}  80\% & \cellcolor{green!70}  76\% & \cellcolor{green!90}  95\% &
\cellcolor{green!100}100\% & \cellcolor{green!100}100\% & \cellcolor{green!100}100\% \\
PF\_02 Reuse test &
\cellcolor{green!80}  84\% & \cellcolor{green!80}  80\% & \cellcolor{green!95}  95\% &
\cellcolor{green!80}  84\% & \cellcolor{green!80}  84\% & \cellcolor{green!100}100\% &
\cellcolor{green!80}  88\% & \cellcolor{green!80}  80\% & \cellcolor{green!90}  91\% &
\cellcolor{green!100}100\% & \cellcolor{green!100}100\% & \cellcolor{green!100}100\% \\
PF\_03 Scale early &
\cellcolor{green!10}  12\% & \cellcolor{green!10}  12\% & \cellcolor{green!100}100\% &
\cellcolor{green!10}  12\% & \cellcolor{green!10}  12\% & \cellcolor{green!100}100\% &
\cellcolor{green!00}  00\% & \cellcolor{green!00}  00\% & \cellcolor{green!00}  ---  &
\cellcolor{green!100}100\% & \cellcolor{green!100}100\% & \cellcolor{green!100}100\% \\
PF\_04 Select early &
\cellcolor{green!00}   0\% & \cellcolor{green!00}   0\% & \cellcolor{green!00}  ---  &
\cellcolor{green!10}  12\% & \cellcolor{green!10}  12\% & \cellcolor{green!100}100\% &
\cellcolor{green!00}   0\% & \cellcolor{green!00}   0\% & \cellcolor{green!00}  ---  &
\cellcolor{green!00}   0\% & \cellcolor{green!00}   0\% & \cellcolor{green!00}  ---  \\
PF\_05 Augment early &
\cellcolor{green!00}   8\% & \cellcolor{green!00}   4\% & \cellcolor{green!50}  50\% &
\cellcolor{green!00}   0\% & \cellcolor{green!00}   0\% & \cellcolor{green!00}  ---  &
\cellcolor{green!00}   4\% & \cellcolor{green!00}   4\% & \cellcolor{green!100}100\% &
\cellcolor{green!00}   0\% & \cellcolor{green!00}   0\% & \cellcolor{green!00}  ---  \\
PF\_06 Subject overlap &
\cellcolor{green!00}   8\% & \cellcolor{green!00}   0\% & \cellcolor{green!00}   0\% &
\cellcolor{green!00}   0\% & \cellcolor{green!00}   0\% & \cellcolor{green!00}  ---  &
\cellcolor{green!00}   0\% & \cellcolor{green!00}   0\% & \cellcolor{green!00}  ---  &
\cellcolor{green!00}   0\% & \cellcolor{green!00}   0\% & \cellcolor{green!00}  ---  \\
PF\_07 Time series &
\cellcolor{green!40}  44\% & \cellcolor{green!40}  40\% & \cellcolor{green!90}  91\% &
\cellcolor{green!20}  24\% & \cellcolor{green!20}  20\% & \cellcolor{green!80}  83\% &
\cellcolor{green!00}   8\% & \cellcolor{green!00}   4\% & \cellcolor{green!50}  50\% &
\cellcolor{green!100}100\% & \cellcolor{green!100}100\% & \cellcolor{green!100}100\% \\
PF\_08 HPO no nest &
\cellcolor{green!00}   8\% & \cellcolor{green!00}   0\% & \cellcolor{green!00}   0\% &
\cellcolor{green!00}   8\% & \cellcolor{green!00}   0\% & \cellcolor{green!00}   0\% &
\cellcolor{green!00}   8\% & \cellcolor{green!00}   0\% & \cellcolor{green!00}   0\% &
\cellcolor{green!00}   0\% & \cellcolor{green!00}   0\% & \cellcolor{green!00}  ---  \\
PF\_09 Ensemble leak &
\cellcolor{green!60}  68\% & \cellcolor{green!20}  24\% & \cellcolor{green!30}  35\% &
\cellcolor{green!40}  40\% & \cellcolor{green!00}   8\% & \cellcolor{green!20}  20\% &
\cellcolor{green!60}  68\% & \cellcolor{green!10}  16\% & \cellcolor{green!20}  24\% &
\cellcolor{green!100}100\% & \cellcolor{green!100}100\% & \cellcolor{green!100}100\% \\
PF\_10 Embed leak &
\cellcolor{green!40}  44\% & \cellcolor{green!00}   8\% & \cellcolor{green!10}  18\% &
\cellcolor{green!10}  16\% & \cellcolor{green!00}   8\% & \cellcolor{green!50}  50\% &
\cellcolor{green!20}  28\% & \cellcolor{green!20}  20\% & \cellcolor{green!70}  71\% &
\cellcolor{green!100}100\% & \cellcolor{green!100}100\% & \cellcolor{green!100}100\% \\
PF\_11 Foundation leak &
\cellcolor{green!00}   0\% & \cellcolor{green!00}   0\% & \cellcolor{green!00}  ---  &
\cellcolor{green!00}   0\% & \cellcolor{green!00}   0\% & \cellcolor{green!00}  ---  &
\cellcolor{green!00}   0\% & \cellcolor{green!00}   0\% & \cellcolor{green!00}  ---  &
\cellcolor{green!00}   0\% & \cellcolor{green!00}   0\% & \cellcolor{green!00}  ---  \\
PF\_12 Bad features &
\cellcolor{green!50}  56\% & \cellcolor{green!50}  56\% & \cellcolor{green!100}100\% &
\cellcolor{green!40}  48\% & \cellcolor{green!40}  48\% & \cellcolor{green!100}100\% &
\cellcolor{green!20}  20\% & \cellcolor{green!20}  20\% & \cellcolor{green!100}100\% &
\cellcolor{green!00}   0\% & \cellcolor{green!00}   0\% & \cellcolor{green!00}  ---  \\
PF\_13 Wrong task &
\cellcolor{green!20}  28\% & \cellcolor{green!20}  28\% & \cellcolor{green!100}100\% &
\cellcolor{green!40}  44\% & \cellcolor{green!40}  44\% & \cellcolor{green!100}100\% &
\cellcolor{green!30}  32\% & \cellcolor{green!30}  32\% & \cellcolor{green!100}100\% &
\cellcolor{green!100}100\% & \cellcolor{green!100}100\% & \cellcolor{green!100}100\% \\
PF\_14 No HPO &
\cellcolor{green!70}  72\% & \cellcolor{green!60}  64\% & \cellcolor{green!80}  89\% &
\cellcolor{green!80}  84\% & \cellcolor{green!60}  60\% & \cellcolor{green!70}  71\% &
\cellcolor{green!80}  80\% & \cellcolor{green!60}  60\% & \cellcolor{green!70}  75\% &
\cellcolor{green!00}   0\% & \cellcolor{green!00}   0\% & \cellcolor{green!00}  ---  \\
PF\_15 No regularize &
\cellcolor{green!20}  20\% & \cellcolor{green!20}  20\% & \cellcolor{green!100}100\% &
\cellcolor{green!70}  72\% & \cellcolor{green!70}  72\% & \cellcolor{green!100}100\% &
\cellcolor{green!10}  16\% & \cellcolor{green!10}  16\% & \cellcolor{green!100}100\% &
\cellcolor{green!00}   0\% & \cellcolor{green!00}   0\% & \cellcolor{green!00}  ---  \\
PF\_16 Using accuracy &
\cellcolor{green!90}  92\% & \cellcolor{green!90}  92\% & \cellcolor{green!100}100\% &
\cellcolor{green!90}  96\% & \cellcolor{green!90}  96\% & \cellcolor{green!100}100\% &
\cellcolor{green!70}  72\% & \cellcolor{green!70}  72\% & \cellcolor{green!100}100\% &
\cellcolor{green!100}100\% & \cellcolor{green!100}100\% & \cellcolor{green!100}100\% \\
PF\_17 Ill-suited model &
\cellcolor{green!00}   4\% & \cellcolor{green!00}   0\% & \cellcolor{green!00}   0\% &
\cellcolor{green!00}   4\% & \cellcolor{green!00}   0\% & \cellcolor{green!00}   0\% &
\cellcolor{green!10}  16\% & \cellcolor{green!00}   0\% & \cellcolor{green!00}   0\% &
\cellcolor{green!100}100\% & \cellcolor{green!100}100\% & \cellcolor{green!100}100\% \\
PF\_18 No baseline &
\cellcolor{green!00}   0\% & \cellcolor{green!00}   0\% & \cellcolor{green!00}  ---  &
\cellcolor{green!00}   0\% & \cellcolor{green!00}   0\% & \cellcolor{green!00}  ---  &
\cellcolor{green!00}   0\% & \cellcolor{green!00}   0\% & \cellcolor{green!00}  ---  &
\cellcolor{green!00}   0\% & \cellcolor{green!00}   0\% & \cellcolor{green!00}  ---  \\
PF\_19 Impute early &
\cellcolor{green!10}  12\% & \cellcolor{green!10}  12\% & \cellcolor{green!100}100\% &
\cellcolor{green!00}   0\% & \cellcolor{green!00}   0\% & \cellcolor{green!00}  ---  &
\cellcolor{green!00}   0\% & \cellcolor{green!00}   0\% & \cellcolor{green!00}  ---  &
\cellcolor{green!00}   0\% & \cellcolor{green!00}   0\% & \cellcolor{green!00}  ---  \\
PF\_20 One metric &
\cellcolor{green!80}  88\% & \cellcolor{green!80}  84\% & \cellcolor{green!90}  95\% &
\cellcolor{green!90}  96\% & \cellcolor{green!90}  96\% & \cellcolor{green!100}100\% &
\cellcolor{green!70}  72\% & \cellcolor{green!60}  60\% & \cellcolor{green!80}  83\% &
\cellcolor{green!100}100\% & \cellcolor{green!100}100\% & \cellcolor{green!100}100\% \\
\bottomrule
\end{tabular}
}
\caption{KM and KH type feedback in LLM responses. The final column for each model shows which proportion of responses with KM also include KH, or ``---'' where KM is zero.}
\label{tab:kmkh}
\end{table}

\section {Limitations} \label{sec:limitations}

This is primarily a study of whether LLMs are able to identify common pitfalls in student code, and whether they can generate useful feedback to learners without additional fine-tuning or prompt engineering. The main limitations are (i) we do not consider the ability of LLMs to recognise the absence of errors in code or the presence of multiple errors at once, which could also be of interest from a pedagogical perspective; (ii) we do not consider the influence of prompt engineering, which in practice could be used to better focus and structure feedback towards learners; (iii) the set of pitfalls, whilst representative of common errors, is not exhaustive; (iv) to make the study manageable in terms of assessing and coding responses, we only consider one code sample per pitfall; (v) although they appear typical of student code, the code samples are semi-synthetic, so there could be bias towards particular LLM-generated styles; (vi) this study only examines the usability of feedback from an educator's perspective, rather than a learner's viewpoint --- this would be especially worth considering in future work, to better understand how LLMs can be best utilised within the learning process.

\section{Conclusions} \label{sec:conclusions}

In this paper, we evaluated the ability of LLMs to both identify common ML pitfalls within code and generate feedback that is suitable for learners. We compared four LLMs: three open models, and one closed. The main finding is that, across all the pitfalls we considered, none of the LLMs were capable of identifying issues within code more than 50\% of the time. In general, this suggests that LLMs are not currently at a level of capability where they can reliably support ML education.

Another prominent concern is the limited degree of overlap between the issues that commonly occur in ML practice and those which LLMs can readily identify. They perform well at identifying the kind of errors that students make early in their ML education, such as not using test sets or using accuracy with imbalanced datasets, but are much less able to spot the kind of issues that occur when ML pipelines increase in complexity. For example, data leaks and overfitting due to the incorrect use of feature selection, data augmentation, hyperparameter optimisation, and foundation models were largely missed by all the LLMs, yet these are common elements of modern ML pipelines. Also notable is their failure to pick up issues at earlier stages of the ML pipeline, where the consequences of a mistake can be more impactful over the course of a project. For example, three of the four models could not identify the error of carrying out mean imputation on the entire dataset before splitting off a test set, something that could lead to an incorrect assessment of model generality.

However, a positive observation is that LLMs do, in general, generate feedback that is beneficial to learners. In the majority of cases, the feedback generated contained useful information about both the issues in the code and information about how to remedy them, both of which are important within an educational context. It seems likely that with further prompt optimisation, the feedback generated by LLMs could be shaped to meet educational needs. For example, to prevent overreliance and promote students’ abilities to code independently, LLMs could be asked to not generate replacement code that a student could directly copy-and-paste into their work. Instead, it could be prompted to focus on forms of feedback which are more valuable for learners.

Different LLMs were able to identify similar pitfalls, but there were notable differences in their success rates at doing so. Whilst the overall mean recall was higher for the closed Github Copilot model, this hides the fact that the open models were more variable in their responses, and were sometimes able to spot errors that the closed model missed, albeit not every time they were prompted. This presumably has a lot to do with the low default temperature setting of Copilot, which is not available for manipulation by users, and appears to strongly discourage non-deterministic behaviour. Whilst this has the potential benefit that the LLM will always produce the same response to the same prompt, it does mean that it was unable to produce any correct responses for 50\% of the pitfalls. By comparison, DeepSeek Coder (one of the open models) was able to produce some correct responses to 85\% of the pitfalls.

One of the main differences between Copilot and the open models we considered is their size, in terms of parameter count, with most estimates placing the former to be at least an order of magnitude larger in size than the latter. Given this large difference in size, the observed differences in capabilities were relatively small. This suggests that there could be a role for using open models in ML education. Unlike closed models, these can be freely deployed on local servers, and do not expose student data beyond the institution. They also offer ease of configurability, with the possibility of fine-tuning their behaviour on locally-collected data. For instance, samples of student work could be collected by an institution, and used to train and configure an open LLM for use within a specific learning context.

Overall, our findings suggest that LLMs, at least in their pretrained state, are not ready to be used within more advanced ML courses and during the transition to practice. This may of course change, and it is quite likely that their capabilities will improve to some extent as LLMs continue to develop. However, a factor underlying their limitations is that these kind of errors are commonplace within code written by humans, meaning that many of the code samples on the internet that LLMs were trained on are also likely to contain them. Given the data dependency of current LLMs, this means that pitfalls are somewhat baked into their behaviour, making their recognition and avoidance a difficult barrier to overcome. It is possible that this could be addressed through more careful data selection when training LLMs, though is innately challenging to do when working with very large datasets. Another direction worth exploring would be to employ in-context learning, where an LLM is provided with relevant information at inference time. For instance, when querying the presence of a particular pitfall within a code sample, the LLM prompt could also be provided with contextual information, either in the form of general guidance or specific examples of correct and incorrect code.

\section*{Declarations}

\begin{itemize}
\item Conflict of interest/Competing interests: The authors declare no competing interests.
\item Data availability: The source code and the results are available in the fol-
lowing GitHub repository: \url{https://github.com/ssk705/ML_Pitfalls} The repository is password protected to prevent its usage in training large language models. 
\item Author contribution: All authors were involved in formulating the research questions, designing the study, and writing the manuscript. The first author of this paper conducted the evaluation. The second and third authors were involved in formulating the ground truth, and compiling the results. All authors have reviewed and approved the final manuscript.
\end{itemize}

\noindent
  

\begin{thebibliography}{}
\renewcommand{\doi}[1]{\url{https://doi.org/#1}}
\bibcommenthead

\bibitem [\protect \citeauthoryear {%
Austin%
\ \protect \BOthers {.}}{%
Austin%
\ \protect \BOthers {.}}{%
{\protect \APACyear {2021}}%
}]{%
austin2021programsynthesislargelanguage}
\APACinsertmetastar {%
austin2021programsynthesislargelanguage}%
\begin{APACrefauthors}%
Austin, J.%
, Odena, A.%
, Nye, M.%
, Bosma, M.%
, Michalewski, H.%
, Dohan, D.%
\BDBL {}Sutton, C.%
\end{APACrefauthors}%
\unskip\
\newblock
\APACrefYearMonthDay{2021}{}{}.
\newblock
\APACrefbtitle {Program Synthesis with Large Language Models.} {Program synthesis with large language models.}
\newblock
\begin{APACrefURL} {https://arxiv.org/abs/2108.07732} \end{APACrefURL}
\PrintBackRefs{\CurrentBib}

\bibitem [\protect \citeauthoryear {%
Becker%
\ \protect \BOthers {.}}{%
Becker%
\ \protect \BOthers {.}}{%
{\protect \APACyear {2023}}%
}]{%
10.1145/3545945.3569759}
\APACinsertmetastar {%
10.1145/3545945.3569759}%
\begin{APACrefauthors}%
Becker, B.A.%
, Denny, P.%
, Finnie-Ansley, J.%
, Luxton-Reilly, A.%
, Prather, J.%
\BCBL {} Santos, E.A.%
\end{APACrefauthors}%
\unskip\
\newblock
\APACrefYearMonthDay{2023}{}{}.
\newblock
{\BBOQ}\APACrefatitle {Programming Is Hard - Or at Least It Used to Be: Educational Opportunities and Challenges of AI Code Generation} {Programming is hard - or at least it used to be: Educational opportunities and challenges of ai code generation}.{\BBCQ}
\newblock
 \APACrefbtitle {Proceedings of the 54th ACM Technical Symposium on Computer Science Education V. 1} {Proceedings of the 54th acm technical symposium on computer science education v. 1}\ (\BPG~500–506).
\newblock
\APACaddressPublisher{New York, NY, USA}{Association for Computing Machinery}.
\newblock
\begin{APACrefURL} {https://doi.org/10.1145/3545945.3569759} \end{APACrefURL}
\PrintBackRefs{\CurrentBib}

\bibitem [\protect \citeauthoryear {%
BigCode%
}{%
BigCode%
}{%
{\protect \APACyear {2024}}%
}]{%
huggingfaceBigcodestarcoderHugging}
\APACinsertmetastar {%
huggingfaceBigcodestarcoderHugging}%
\begin{APACrefauthors}%
BigCode%
\end{APACrefauthors}%
\unskip\
\newblock
\APACrefYearMonthDay{2024}{}{}.
\newblock
\APACrefbtitle {bigcode/starcoder · {H}ugging {F}ace --- huggingface.co.} {bigcode/starcoder · {H}ugging {F}ace --- huggingface.co.}
\newblock
\APAChowpublished {\url{https://huggingface.co/bigcode/starcoder}}.
\newblock
\APACrefnote{[Accessed 03-11-2024]}
\PrintBackRefs{\CurrentBib}

\bibitem [\protect \citeauthoryear {%
Burtsev%
, Reeves%
\BCBL {}\ \BBA {} Job%
}{%
Burtsev%
\ \protect \BOthers {.}}{%
{\protect \APACyear {2024}}%
}]{%
burtsev2024working}
\APACinsertmetastar {%
burtsev2024working}%
\begin{APACrefauthors}%
Burtsev, M.%
, Reeves, M.%
\BCBL {} Job, A.%
\end{APACrefauthors}%
\unskip\
\newblock
\APACrefYearMonthDay{2024}{}{}.
\newblock
{\BBOQ}\APACrefatitle {The working limitations of large language models} {The working limitations of large language models}.{\BBCQ}
\newblock
\APACjournalVolNumPages{MIT Sloan Management Review}{65}{2}{8--10,}
\newblock
\begin{APACrefURL} {https://sloanreview.mit.edu/article/the-working-limitations-of-large-language-models/} \end{APACrefURL}
\newblock

\newblock

\PrintBackRefs{\CurrentBib}

\bibitem [\protect \citeauthoryear {%
Cambaz%
\ \BBA {} Zhang%
}{%
Cambaz%
\ \BBA {} Zhang%
}{%
{\protect \APACyear {2024}}%
}]{%
10.1145/3626252.3630958}
\APACinsertmetastar {%
10.1145/3626252.3630958}%
\begin{APACrefauthors}%
Cambaz, D.%
\BCBT {}\ \BBA {} Zhang, X.%
\end{APACrefauthors}%
\unskip\
\newblock
\APACrefYearMonthDay{2024}{}{}.
\newblock
{\BBOQ}\APACrefatitle {Use of AI-driven Code Generation Models in Teaching and Learning Programming: a Systematic Literature Review} {Use of ai-driven code generation models in teaching and learning programming: a systematic literature review}.{\BBCQ}
\newblock
 \APACrefbtitle {Proceedings of the 55th ACM Technical Symposium on Computer Science Education V. 1} {Proceedings of the 55th acm technical symposium on computer science education v. 1}\ (\BPG~172–178).
\newblock
\APACaddressPublisher{New York, NY, USA}{Association for Computing Machinery}.
\newblock
\begin{APACrefURL} {https://doi.org/10.1145/3626252.3630958} \end{APACrefURL}
\PrintBackRefs{\CurrentBib}

\bibitem [\protect \citeauthoryear {%
Cheng%
, Li%
\BCBL {}\ \BBA {} Bing%
}{%
Cheng%
\ \protect \BOthers {.}}{%
{\protect \APACyear {2023}}%
}]{%
cheng2023gpt}
\APACinsertmetastar {%
cheng2023gpt}%
\begin{APACrefauthors}%
Cheng, L.%
, Li, X.%
\BCBL {} Bing, L.%
\end{APACrefauthors}%
\unskip\
\newblock
\APACrefYearMonthDay{2023}{}{}.
\newblock
{\BBOQ}\APACrefatitle {Is {GPT-4} a Good Data Analyst?} {Is {GPT-4} a good data analyst?}{\BBCQ}
\newblock
 H.~Bouamor, J.~Pino\BCBL {}\ \BBA {} K.~Bali\ (\BEDS), \APACrefbtitle {Findings of the Association for Computational Linguistics: {EMNLP} 2023, Singapore, December 6-10, 2023} {Findings of the association for computational linguistics: {EMNLP} 2023, singapore, december 6-10, 2023}\ (\BPGS\ 9496--9514).
\newblock
\APACaddressPublisher{}{Association for Computational Linguistics}.
\newblock
\begin{APACrefURL} {https://doi.org/10.18653/v1/2023.findings-emnlp.637} \end{APACrefURL}
\PrintBackRefs{\CurrentBib}

\bibitem [\protect \citeauthoryear {%
Cinca%
, Costanza%
, Musolesi%
\BCBL {}\ \BBA {} Alebri%
}{%
Cinca%
\ \protect \BOthers {.}}{%
{\protect \APACyear {2025}}%
}]{%
cinca2025they}
\APACinsertmetastar {%
cinca2025they}%
\begin{APACrefauthors}%
Cinca, R.%
, Costanza, E.%
, Musolesi, M.%
\BCBL {} Alebri, M.%
\end{APACrefauthors}%
\unskip\
\newblock
\APACrefYearMonthDay{2025}{}{}.
\newblock
{\BBOQ}\APACrefatitle {“What are they not telling me?” Learning machine learning: Understanding the challenges for novices} {“what are they not telling me?” learning machine learning: Understanding the challenges for novices}.{\BBCQ}
\newblock
\APACjournalVolNumPages{International Journal of Human-Computer Studies}{196}{}{103438,}
\newblock
\begin{APACrefDOI} \doi{10.1016/j.ijhcs.2024.103438} \end{APACrefDOI}
\newblock

\newblock

\PrintBackRefs{\CurrentBib}

\bibitem [\protect \citeauthoryear {%
CodeLlama%
}{%
CodeLlama%
}{%
{\protect \APACyear {2024}}%
}]{%
codeLlama}
\APACinsertmetastar {%
codeLlama}%
\begin{APACrefauthors}%
CodeLlama%
\end{APACrefauthors}%
\unskip\
\newblock
\APACrefYearMonthDay{2024}{}{}.
\newblock
\APACrefbtitle {codellama/{C}ode{L}lama-34b-{I}nstruct-hf · {H}ugging {F}ace --- huggingface.co.} {codellama/{C}ode{L}lama-34b-{I}nstruct-hf · {H}ugging {F}ace --- huggingface.co.}
\newblock
\APAChowpublished {\url{https://huggingface.co/codellama/CodeLlama-34b-Instruct-hf}}.
\newblock
\APACrefnote{[Accessed 26-10-2024]}
\PrintBackRefs{\CurrentBib}

\bibitem [\protect \citeauthoryear {%
Cooper%
}{%
Cooper%
}{%
{\protect \APACyear {2024}}%
}]{%
cooper2024ai}
\APACinsertmetastar {%
cooper2024ai}%
\begin{APACrefauthors}%
Cooper, R.G.%
\end{APACrefauthors}%
\unskip\
\newblock
\APACrefYearMonthDay{2024}{}{}.
\newblock
{\BBOQ}\APACrefatitle {Why {AI} projects fail: Lessons from new product development} {Why {AI} projects fail: Lessons from new product development}.{\BBCQ}
\newblock
\APACjournalVolNumPages{IEEE Engineering Management Review}{52}{4}{15--21,}
\newblock
\begin{APACrefDOI} \doi{10.1109/EMR.2024.3419268} \end{APACrefDOI}
\newblock

\newblock

\PrintBackRefs{\CurrentBib}

\bibitem [\protect \citeauthoryear {%
DeepSeek%
}{%
DeepSeek%
}{%
{\protect \APACyear {2024}}%
}]{%
deepseek}
\APACinsertmetastar {%
deepseek}%
\begin{APACrefauthors}%
DeepSeek%
\end{APACrefauthors}%
\unskip\
\newblock
\APACrefYearMonthDay{2024}{}{}.
\newblock
\APACrefbtitle {deepseek-ai/{D}eep{S}eek-{C}oder-{V}2-{L}ite-{I}nstruct · {H}ugging {F}ace --- huggingface.co.} {deepseek-ai/{D}eep{S}eek-{C}oder-{V}2-{L}ite-{I}nstruct · {H}ugging {F}ace --- huggingface.co.}
\newblock
\APAChowpublished {\url{https://huggingface.co/deepseek-ai/DeepSeek-Coder-V2-Lite-Instruct}}.
\newblock
\APACrefnote{[Accessed 26-10-2024]}
\PrintBackRefs{\CurrentBib}

\bibitem [\protect \citeauthoryear {%
Denny%
, Luxton-Reilly%
\BCBL {}\ \BBA {} Tempero%
}{%
Denny%
\ \protect \BOthers {.}}{%
{\protect \APACyear {2012}}%
}]{%
denny2012all}
\APACinsertmetastar {%
denny2012all}%
\begin{APACrefauthors}%
Denny, P.%
, Luxton-Reilly, A.%
\BCBL {} Tempero, E.%
\end{APACrefauthors}%
\unskip\
\newblock
\APACrefYearMonthDay{2012}{}{}.
\newblock
{\BBOQ}\APACrefatitle {All syntax errors are not equal} {All syntax errors are not equal}.{\BBCQ}
\newblock
 \APACrefbtitle {Proceedings of the 17th ACM annual conference on Innovation and Technology in Computer Science Education} {Proceedings of the 17th acm annual conference on innovation and technology in computer science education}\ (\BPGS\ 75--80).
\newblock
\begin{APACrefURL} {https://doi.org/10.1145/2325296.2325318} \end{APACrefURL}
\PrintBackRefs{\CurrentBib}

\bibitem [\protect \citeauthoryear {%
Denny%
, MacNeil%
, Savelka%
, Porter%
\BCBL {}\ \BBA {} Luxton-Reilly%
}{%
Denny%
\ \protect \BOthers {.}}{%
{\protect \APACyear {2024}}%
}]{%
10.1145/3649217.3653574}
\APACinsertmetastar {%
10.1145/3649217.3653574}%
\begin{APACrefauthors}%
Denny, P.%
, MacNeil, S.%
, Savelka, J.%
, Porter, L.%
\BCBL {} Luxton-Reilly, A.%
\end{APACrefauthors}%
\unskip\
\newblock
\APACrefYearMonthDay{2024}{}{}.
\newblock
{\BBOQ}\APACrefatitle {Desirable Characteristics for AI Teaching Assistants in Programming Education} {Desirable characteristics for ai teaching assistants in programming education}.{\BBCQ}
\newblock
 \APACrefbtitle {Proceedings of the 2024 on Innovation and Technology in Computer Science Education V. 1} {Proceedings of the 2024 on innovation and technology in computer science education v. 1}\ (\BPG~408–414).
\newblock
\APACaddressPublisher{New York, NY, USA}{Association for Computing Machinery}.
\newblock
\begin{APACrefURL} {https://doi.org/10.1145/3649217.3653574} \end{APACrefURL}
\PrintBackRefs{\CurrentBib}

\bibitem [\protect \citeauthoryear {%
Ermakova%
\ \protect \BOthers {.}}{%
Ermakova%
\ \protect \BOthers {.}}{%
{\protect \APACyear {2021}}%
}]{%
ermakova2021beyond}
\APACinsertmetastar {%
ermakova2021beyond}%
\begin{APACrefauthors}%
Ermakova, T.%
, Blume, J.%
, Fabian, B.%
, Fomenko, E.%
, Berlin, M.%
\BCBL {} Hauswirth, M.%
\end{APACrefauthors}%
\unskip\
\newblock
\APACrefYearMonthDay{2021}{}{}.
\newblock
{\BBOQ}\APACrefatitle {Beyond the Hype: Why Do Data-Driven Projects Fail?} {Beyond the hype: Why do data-driven projects fail?}{\BBCQ}
\newblock
 \APACrefbtitle {54th Hawaii International Conference on System Sciences, {HICSS}} {54th hawaii international conference on system sciences, {HICSS}}\ (\BPGS\ 1--10).
\newblock
\APACaddressPublisher{}{ScholarSpace}.
\newblock
\begin{APACrefURL} {https://hdl.handle.net/10125/71237} \end{APACrefURL}
\PrintBackRefs{\CurrentBib}

\bibitem [\protect \citeauthoryear {%
Ettles%
, Luxton-Reilly%
\BCBL {}\ \BBA {} Denny%
}{%
Ettles%
\ \protect \BOthers {.}}{%
{\protect \APACyear {2018}}%
}]{%
ettles2018common}
\APACinsertmetastar {%
ettles2018common}%
\begin{APACrefauthors}%
Ettles, A.%
, Luxton-Reilly, A.%
\BCBL {} Denny, P.%
\end{APACrefauthors}%
\unskip\
\newblock
\APACrefYearMonthDay{2018}{}{}.
\newblock
{\BBOQ}\APACrefatitle {Common logic errors made by novice programmers} {Common logic errors made by novice programmers}.{\BBCQ}
\newblock
 \APACrefbtitle {Proceedings of the 20th Australasian Computing Education Conference} {Proceedings of the 20th australasian computing education conference}\ (\BPGS\ 83--89).
\newblock
\begin{APACrefURL} {https://doi.org/10.1145/3160489.3160493} \end{APACrefURL}
\PrintBackRefs{\CurrentBib}

\bibitem [\protect \citeauthoryear {%
GitHub%
}{%
GitHub%
}{%
{\protect \APACyear {2024}}%
}]{%
githubGitHubCopilot}
\APACinsertmetastar {%
githubGitHubCopilot}%
\begin{APACrefauthors}%
GitHub%
\end{APACrefauthors}%
\unskip\
\newblock
\APACrefYearMonthDay{2024}{}{}.
\newblock
\APACrefbtitle {{G}it{H}ub {C}opilot · {Y}our {A}{I} pair programmer --- github.com.} {{G}it{H}ub {C}opilot · {Y}our {A}{I} pair programmer --- github.com.}
\newblock
\APAChowpublished {\url{https://github.com/features/copilot}}.
\newblock
\APACrefnote{[Accessed 20-11-2024]}
\PrintBackRefs{\CurrentBib}

\bibitem [\protect \citeauthoryear {%
Goldman%
\ \protect \BOthers {.}}{%
Goldman%
\ \protect \BOthers {.}}{%
{\protect \APACyear {2008}}%
}]{%
goldman2008identifying}
\APACinsertmetastar {%
goldman2008identifying}%
\begin{APACrefauthors}%
Goldman, K.%
, Gross, P.%
, Heeren, C.%
, Herman, G.%
, Kaczmarczyk, L.%
, Loui, M.C.%
\BCBL {} Zilles, C.%
\end{APACrefauthors}%
\unskip\
\newblock
\APACrefYearMonthDay{2008}{}{}.
\newblock
{\BBOQ}\APACrefatitle {Identifying important and difficult concepts in introductory computing courses using a delphi process} {Identifying important and difficult concepts in introductory computing courses using a delphi process}.{\BBCQ}
\newblock
 \APACrefbtitle {Proceedings of the 39th SIGCSE technical symposium on Computer science education} {Proceedings of the 39th sigcse technical symposium on computer science education}\ (\BPGS\ 256--260).
\newblock
\begin{APACrefURL} {https://doi.org/10.1145/1352322.1352226} \end{APACrefURL}
\PrintBackRefs{\CurrentBib}

\bibitem [\protect \citeauthoryear {%
Gramoli%
\ \protect \BOthers {.}}{%
Gramoli%
\ \protect \BOthers {.}}{%
{\protect \APACyear {2016}}%
}]{%
10.1145/2843043.2843070}
\APACinsertmetastar {%
10.1145/2843043.2843070}%
\begin{APACrefauthors}%
Gramoli, V.%
, Charleston, M.%
, Jeffries, B.%
, Koprinska, I.%
, McGrane, M.%
, Radu, A.%
\BDBL {}Yacef, K.%
\end{APACrefauthors}%
\unskip\
\newblock
\APACrefYearMonthDay{2016}{}{}.
\newblock
{\BBOQ}\APACrefatitle {Mining autograding data in computer science education} {Mining autograding data in computer science education}.{\BBCQ}
\newblock
 \APACrefbtitle {Proceedings of the Australasian Computer Science Week Multiconference.} {Proceedings of the australasian computer science week multiconference.}
\newblock
\APACaddressPublisher{New York, NY, USA}{Association for Computing Machinery}.
\newblock
\begin{APACrefURL} {https://doi.org/10.1145/2843043.2843070} \end{APACrefURL}
\PrintBackRefs{\CurrentBib}

\bibitem [\protect \citeauthoryear {%
Gulwani%
, Radi{\v{c}}ek%
\BCBL {}\ \BBA {} Zuleger%
}{%
Gulwani%
\ \protect \BOthers {.}}{%
{\protect \APACyear {2018}}%
}]{%
gulwani2018automated}
\APACinsertmetastar {%
gulwani2018automated}%
\begin{APACrefauthors}%
Gulwani, S.%
, Radi{\v{c}}ek, I.%
\BCBL {} Zuleger, F.%
\end{APACrefauthors}%
\unskip\
\newblock
\APACrefYearMonthDay{2018}{}{}.
\newblock
{\BBOQ}\APACrefatitle {Automated clustering and program repair for introductory programming assignments} {Automated clustering and program repair for introductory programming assignments}.{\BBCQ}
\newblock
\APACjournalVolNumPages{ACM SIGPLAN Notices}{53}{4}{465--480,}
\newblock
\begin{APACrefDOI} \doi{10.1145/3192366.3192387} \end{APACrefDOI}
\newblock

\newblock

\PrintBackRefs{\CurrentBib}

\bibitem [\protect \citeauthoryear {%
Hellas%
\ \protect \BOthers {.}}{%
Hellas%
\ \protect \BOthers {.}}{%
{\protect \APACyear {2023}}%
}]{%
hellas2023exploring}
\APACinsertmetastar {%
hellas2023exploring}%
\begin{APACrefauthors}%
Hellas, A.%
, Leinonen, J.%
, Sarsa, S.%
, Koutcheme, C.%
, Kujanp{\"{a}}{\"{a}}, L.%
\BCBL {} Sorva, J.%
\end{APACrefauthors}%
\unskip\
\newblock
\APACrefYearMonthDay{2023}{}{}.
\newblock
{\BBOQ}\APACrefatitle {Exploring the Responses of Large Language Models to Beginner Programmers' Help Requests} {Exploring the responses of large language models to beginner programmers' help requests}.{\BBCQ}
\newblock
 K.~Fisler, P.~Denny, D.~Franklin\BCBL {}\ \BBA {} M.~Hamilton\ (\BEDS), \APACrefbtitle {Proceedings of the 2023 {ACM} Conference on International Computing Education Research - Volume 1, {ICER} 2023, Chicago, IL, USA, August 7-11, 2023} {Proceedings of the 2023 {ACM} conference on international computing education research - volume 1, {ICER} 2023, chicago, il, usa, august 7-11, 2023}\ (\BPGS\ 93--105).
\newblock
\begin{APACrefURL} {https://doi.org/10.1145/3568813.3600139} \end{APACrefURL}
\PrintBackRefs{\CurrentBib}

\bibitem [\protect \citeauthoryear {%
Hewamalage%
, Ackermann%
\BCBL {}\ \BBA {} Bergmeir%
}{%
Hewamalage%
\ \protect \BOthers {.}}{%
{\protect \APACyear {2023}}%
}]{%
hewamalage2023forecast}
\APACinsertmetastar {%
hewamalage2023forecast}%
\begin{APACrefauthors}%
Hewamalage, H.%
, Ackermann, K.%
\BCBL {} Bergmeir, C.%
\end{APACrefauthors}%
\unskip\
\newblock
\APACrefYearMonthDay{2023}{}{}.
\newblock
{\BBOQ}\APACrefatitle {Forecast evaluation for data scientists: common pitfalls and best practices} {Forecast evaluation for data scientists: common pitfalls and best practices}.{\BBCQ}
\newblock
\APACjournalVolNumPages{Data Min. Knowl. Discov.}{37}{2}{788--832,}
\newblock
\begin{APACrefDOI} \doi{10.1007/S10618-022-00894-5} \end{APACrefDOI}
\newblock

\newblock

\PrintBackRefs{\CurrentBib}

\bibitem [\protect \citeauthoryear {%
{Hugging Face}%
}{%
{Hugging Face}%
}{%
{\protect \APACyear {2024}}%
}]{%
huggingfaceCodeModels}
\APACinsertmetastar {%
huggingfaceCodeModels}%
\begin{APACrefauthors}%
{Hugging Face}%
\end{APACrefauthors}%
\unskip\
\newblock
\APACrefYearMonthDay{2024}{}{}.
\newblock
\APACrefbtitle {{B}ig {C}ode {M}odels {L}eaderboard - a {H}ugging {F}ace {S}pace by bigcode --- huggingface.co.} {{B}ig {C}ode {M}odels {L}eaderboard - a {H}ugging {F}ace {S}pace by bigcode --- huggingface.co.}
\newblock
\APAChowpublished {\url{https://huggingface.co/spaces/bigcode/bigcode-models-leaderboard?spm=a2c65.11461447.0.0.4d297330G0fpz7 }}.
\newblock
\APACrefnote{[Accessed 26-10-2024]}
\PrintBackRefs{\CurrentBib}

\bibitem [\protect \citeauthoryear {%
Jiang%
, Wang%
, Shen%
, Kim%
\BCBL {}\ \BBA {} Kim%
}{%
Jiang%
\ \protect \BOthers {.}}{%
{\protect \APACyear {2024}}%
}]{%
jiang2024surveylargelanguagemodels}
\APACinsertmetastar {%
jiang2024surveylargelanguagemodels}%
\begin{APACrefauthors}%
Jiang, J.%
, Wang, F.%
, Shen, J.%
, Kim, S.%
\BCBL {} Kim, S.%
\end{APACrefauthors}%
\unskip\
\newblock
\APACrefYearMonthDay{2024}{}{}.
\newblock
\APACrefbtitle {A Survey on Large Language Models for Code Generation.} {A survey on large language models for code generation.}
\newblock
\begin{APACrefURL} {https://arxiv.org/abs/2406.00515} \end{APACrefURL}
\PrintBackRefs{\CurrentBib}

\bibitem [\protect \citeauthoryear {%
Kapoor%
\ \protect \BOthers {.}}{%
Kapoor%
\ \protect \BOthers {.}}{%
{\protect \APACyear {2024}}%
}]{%
ref1}
\APACinsertmetastar {%
ref1}%
\begin{APACrefauthors}%
Kapoor, S.%
, Cantrell, E.M.%
, Peng, K.%
, Pham, T.H.%
, Bail, C.A.%
, Gundersen, O.E.%
\BDBL {}Narayanan, A.%
\end{APACrefauthors}%
\unskip\
\newblock
\APACrefYearMonthDay{2024}{2024/10/23}{}.
\newblock
{\BBOQ}\APACrefatitle {REFORMS: Consensus-based Recommendations for Machine-learning-based Science} {Reforms: Consensus-based recommendations for machine-learning-based science}.{\BBCQ}
\newblock
\APACjournalVolNumPages{Science Advances}{10}{18}{eadk3452,}
\newblock
\begin{APACrefDOI} \doi{10.1126/sciadv.adk3452} \end{APACrefDOI}
\newblock

\newblock

\PrintBackRefs{\CurrentBib}

\bibitem [\protect \citeauthoryear {%
Kapoor%
\ \BBA {} Narayanan%
}{%
Kapoor%
\ \BBA {} Narayanan%
}{%
{\protect \APACyear {2023}}%
}]{%
KAPOOR2023100804}
\APACinsertmetastar {%
KAPOOR2023100804}%
\begin{APACrefauthors}%
Kapoor, S.%
\BCBT {}\ \BBA {} Narayanan, A.%
\end{APACrefauthors}%
\unskip\
\newblock
\APACrefYearMonthDay{2023}{}{}.
\newblock
{\BBOQ}\APACrefatitle {Leakage and the reproducibility crisis in machine-learning-based science} {Leakage and the reproducibility crisis in machine-learning-based science}.{\BBCQ}
\newblock
\APACjournalVolNumPages{Patterns}{4}{9}{100804,}
\newblock
\begin{APACrefDOI} \doi{https://doi.org/10.1016/j.patter.2023.100804} \end{APACrefDOI}
\newblock

\newblock

\PrintBackRefs{\CurrentBib}

\bibitem [\protect \citeauthoryear {%
Kasneci%
\ \protect \BOthers {.}}{%
Kasneci%
\ \protect \BOthers {.}}{%
{\protect \APACyear {2023}}%
}]{%
KASNECI2023102274}
\APACinsertmetastar {%
KASNECI2023102274}%
\begin{APACrefauthors}%
Kasneci, E.%
, Sessler, K.%
, Küchemann, S.%
, Bannert, M.%
, Dementieva, D.%
, Fischer, F.%
\BDBL {}Kasneci, G.%
\end{APACrefauthors}%
\unskip\
\newblock
\APACrefYearMonthDay{2023}{}{}.
\newblock
{\BBOQ}\APACrefatitle {ChatGPT for good? On opportunities and challenges of large language models for education} {Chatgpt for good? on opportunities and challenges of large language models for education}.{\BBCQ}
\newblock
\APACjournalVolNumPages{Learning and Individual Differences}{103}{}{102274,}
\newblock
\begin{APACrefDOI} \doi{https://doi.org/10.1016/j.lindif.2023.102274} \end{APACrefDOI}
\newblock

\newblock

\PrintBackRefs{\CurrentBib}

\bibitem [\protect \citeauthoryear {%
Keuning%
, Jeuring%
\BCBL {}\ \BBA {} Heeren%
}{%
Keuning%
\ \protect \BOthers {.}}{%
{\protect \APACyear {2018}}%
}]{%
keuning2018systematic}
\APACinsertmetastar {%
keuning2018systematic}%
\begin{APACrefauthors}%
Keuning, H.%
, Jeuring, J.%
\BCBL {} Heeren, B.%
\end{APACrefauthors}%
\unskip\
\newblock
\APACrefYearMonthDay{2018}{}{}.
\newblock
{\BBOQ}\APACrefatitle {A systematic literature review of automated feedback generation for programming exercises} {A systematic literature review of automated feedback generation for programming exercises}.{\BBCQ}
\newblock
\APACjournalVolNumPages{ACM Transactions on Computing Education (TOCE)}{19}{1}{1--43,}
\newblock
\begin{APACrefDOI} \doi{10.1145/3231711} \end{APACrefDOI}
\newblock

\newblock

\PrintBackRefs{\CurrentBib}

\bibitem [\protect \citeauthoryear {%
Leinonen%
\ \protect \BOthers {.}}{%
Leinonen%
\ \protect \BOthers {.}}{%
{\protect \APACyear {2024}}%
}]{%
leinonen2024llmitationsincerestformdata}
\APACinsertmetastar {%
leinonen2024llmitationsincerestformdata}%
\begin{APACrefauthors}%
Leinonen, J.%
, Denny, P.%
, Kiljunen, O.%
, MacNeil, S.%
, Sarsa, S.%
\BCBL {} Hellas, A.%
\end{APACrefauthors}%
\unskip\
\newblock
\APACrefYearMonthDay{2024}{}{}.
\newblock
\APACrefbtitle {LLM-itation is the Sincerest Form of Data: Generating Synthetic Buggy Code Submissions for Computing Education.} {Llm-itation is the sincerest form of data: Generating synthetic buggy code submissions for computing education.}
\newblock
\begin{APACrefURL} {https://arxiv.org/abs/2411.10455} \end{APACrefURL}
\PrintBackRefs{\CurrentBib}

\bibitem [\protect \citeauthoryear {%
Liao%
, Taori%
, Raji%
\BCBL {}\ \BBA {} Schmidt%
}{%
Liao%
\ \protect \BOthers {.}}{%
{\protect \APACyear {2021}}%
}]{%
liao2021we}
\APACinsertmetastar {%
liao2021we}%
\begin{APACrefauthors}%
Liao, T.%
, Taori, R.%
, Raji, D.%
\BCBL {} Schmidt, L.%
\end{APACrefauthors}%
\unskip\
\newblock
\APACrefYearMonthDay{2021}{}{}.
\newblock
{\BBOQ}\APACrefatitle {Are We Learning Yet? {A} Meta Review of Evaluation Failures Across Machine Learning} {Are we learning yet? {A} meta review of evaluation failures across machine learning}.{\BBCQ}
\newblock
 J.~Vanschoren\ \BBA {} S.~Yeung\ (\BEDS), \APACrefbtitle {Proceedings of the Neural Information Processing Systems Track on Datasets and Benchmarks 1, NeurIPS Datasets and Benchmarks 2021, December 2021, virtual.} {Proceedings of the neural information processing systems track on datasets and benchmarks 1, neurips datasets and benchmarks 2021, december 2021, virtual.}
\newblock
\begin{APACrefURL} {https://datasets-benchmarks-proceedings.neurips.cc/paper/2021/hash/757b505cfd34c64c85ca5b5690ee5293-Abstract-round2.html} \end{APACrefURL}
\PrintBackRefs{\CurrentBib}

\bibitem [\protect \citeauthoryear {%
Liu%
\ \protect \BOthers {.}}{%
Liu%
\ \protect \BOthers {.}}{%
{\protect \APACyear {2024}}%
}]{%
liu2024llmpoweredtestcasegeneration}
\APACinsertmetastar {%
liu2024llmpoweredtestcasegeneration}%
\begin{APACrefauthors}%
Liu, K.%
, Liu, Y.%
, Chen, Z.%
, Zhang, J.M.%
, Han, Y.%
, Ma, Y.%
\BDBL {}Huang, G.%
\end{APACrefauthors}%
\unskip\
\newblock
\APACrefYearMonthDay{2024}{}{}.
\newblock
\APACrefbtitle {LLM-Powered Test Case Generation for Detecting Tricky Bugs.} {Llm-powered test case generation for detecting tricky bugs.}
\newblock
\begin{APACrefURL} {https://arxiv.org/abs/2404.10304} \end{APACrefURL}
\PrintBackRefs{\CurrentBib}

\bibitem [\protect \citeauthoryear {%
Lones%
}{%
Lones%
}{%
{\protect \APACyear {2024}}%
}]{%
Lones2024}
\APACinsertmetastar {%
Lones2024}%
\begin{APACrefauthors}%
Lones, M.A.%
\end{APACrefauthors}%
\unskip\
\newblock
\APACrefYearMonthDay{2024}{Oct}{11}.
\newblock
{\BBOQ}\APACrefatitle {Avoiding common machine learning pitfalls} {Avoiding common machine learning pitfalls}.{\BBCQ}
\newblock
\APACjournalVolNumPages{Patterns}{5}{10}{,}
\newblock
\begin{APACrefDOI} \doi{10.1016/j.patter.2024.101046} \end{APACrefDOI}
\newblock

\newblock

\PrintBackRefs{\CurrentBib}

\bibitem [\protect \citeauthoryear {%
Lyu%
, Wang%
, Chung%
, Sun%
\BCBL {}\ \BBA {} Zhang%
}{%
Lyu%
\ \protect \BOthers {.}}{%
{\protect \APACyear {2024}}%
}]{%
10.1145/3657604.3662036}
\APACinsertmetastar {%
10.1145/3657604.3662036}%
\begin{APACrefauthors}%
Lyu, W.%
, Wang, Y.%
, Chung, T.R.%
, Sun, Y.%
\BCBL {} Zhang, Y.%
\end{APACrefauthors}%
\unskip\
\newblock
\APACrefYearMonthDay{2024}{}{}.
\newblock
{\BBOQ}\APACrefatitle {Evaluating the Effectiveness of LLMs in Introductory Computer Science Education: A Semester-Long Field Study} {Evaluating the effectiveness of llms in introductory computer science education: A semester-long field study}.{\BBCQ}
\newblock
 \APACrefbtitle {Proceedings of the Eleventh ACM Conference on Learning @ Scale} {Proceedings of the eleventh acm conference on learning @ scale}\ (\BPG~63–74).
\newblock
\APACaddressPublisher{New York, NY, USA}{Association for Computing Machinery}.
\newblock
\begin{APACrefURL} {https://doi.org/10.1145/3657604.3662036} \end{APACrefURL}
\PrintBackRefs{\CurrentBib}

\bibitem [\protect \citeauthoryear {%
MacNeil%
\ \protect \BOthers {.}}{%
MacNeil%
\ \protect \BOthers {.}}{%
{\protect \APACyear {2023}}%
}]{%
10.1145/3545945.3569785}
\APACinsertmetastar {%
10.1145/3545945.3569785}%
\begin{APACrefauthors}%
MacNeil, S.%
, Tran, A.%
, Hellas, A.%
, Kim, J.%
, Sarsa, S.%
, Denny, P.%
\BDBL {}Leinonen, J.%
\end{APACrefauthors}%
\unskip\
\newblock
\APACrefYearMonthDay{2023}{}{}.
\newblock
{\BBOQ}\APACrefatitle {Experiences from Using Code Explanations Generated by Large Language Models in a Web Software Development E-Book} {Experiences from using code explanations generated by large language models in a web software development e-book}.{\BBCQ}
\newblock
 \APACrefbtitle {Proceedings of the 54th ACM Technical Symposium on Computer Science Education V. 1} {Proceedings of the 54th acm technical symposium on computer science education v. 1}\ (\BPG~931–937).
\newblock
\APACaddressPublisher{New York, NY, USA}{Association for Computing Machinery}.
\newblock
\begin{APACrefURL} {https://doi.org/10.1145/3545945.3569785} \end{APACrefURL}
\PrintBackRefs{\CurrentBib}

\bibitem [\protect \citeauthoryear {%
Nahar%
, Zhang%
, Lewis%
, Zhou%
\BCBL {}\ \BBA {} K{\"{a}}stner%
}{%
Nahar%
\ \protect \BOthers {.}}{%
{\protect \APACyear {2023}}%
}]{%
nahar2023meta}
\APACinsertmetastar {%
nahar2023meta}%
\begin{APACrefauthors}%
Nahar, N.%
, Zhang, H.%
, Lewis, G.A.%
, Zhou, S.%
\BCBL {} K{\"{a}}stner, C.%
\end{APACrefauthors}%
\unskip\
\newblock
\APACrefYearMonthDay{2023}{}{}.
\newblock
{\BBOQ}\APACrefatitle {A Meta-Summary of Challenges in Building Products with {ML} Components - Collecting Experiences from 4758+ Practitioners} {A meta-summary of challenges in building products with {ML} components - collecting experiences from 4758+ practitioners}.{\BBCQ}
\newblock
 \APACrefbtitle {2nd {IEEE/ACM} International Conference on {AI} Engineering - Software Engineering for AI, {CAIN} 2023, Melbourne, Australia, May 15-16, 2023} {2nd {IEEE/ACM} international conference on {AI} engineering - software engineering for ai, {CAIN} 2023, melbourne, australia, may 15-16, 2023}\ (\BPGS\ 171--183).
\newblock
\begin{APACrefURL} {https://doi.org/10.1109/CAIN58948.2023.00034} \end{APACrefURL}
\PrintBackRefs{\CurrentBib}

\bibitem [\protect \citeauthoryear {%
Narciss%
}{%
Narciss%
}{%
{\protect \APACyear {2008}}%
}]{%
narciss2008feedback}
\APACinsertmetastar {%
narciss2008feedback}%
\begin{APACrefauthors}%
Narciss, S.%
\end{APACrefauthors}%
\unskip\
\newblock
\APACrefYearMonthDay{2008}{}{}.
\newblock
{\BBOQ}\APACrefatitle {Feedback strategies for interactive learning tasks} {Feedback strategies for interactive learning tasks}.{\BBCQ}
\newblock
 \APACrefbtitle {Handbook of research on educational communications and technology} {Handbook of research on educational communications and technology}\ (\BPGS\ 125--143).
\newblock
\APACaddressPublisher{}{Routledge}.
\newblock
\begin{APACrefURL} {https://www.taylorfrancis.com/chapters/edit/10.4324/9780203880869-13/feedback-strategies-interactive-learning-tasks-susanne-narciss} \end{APACrefURL}
\PrintBackRefs{\CurrentBib}

\bibitem [\protect \citeauthoryear {%
Naveed%
\ \protect \BOthers {.}}{%
Naveed%
\ \protect \BOthers {.}}{%
{\protect \APACyear {2024}}%
}]{%
naveed2024comprehensiveoverviewlargelanguage}
\APACinsertmetastar {%
naveed2024comprehensiveoverviewlargelanguage}%
\begin{APACrefauthors}%
Naveed, H.%
, Khan, A.U.%
, Qiu, S.%
, Saqib, M.%
, Anwar, S.%
, Usman, M.%
\BDBL {}Mian, A.%
\end{APACrefauthors}%
\unskip\
\newblock
\APACrefYearMonthDay{2024}{}{}.
\newblock
\APACrefbtitle {A Comprehensive Overview of Large Language Models.} {A comprehensive overview of large language models.}
\newblock
\begin{APACrefURL} {https://arxiv.org/abs/2307.06435} \end{APACrefURL}
\PrintBackRefs{\CurrentBib}

\bibitem [\protect \citeauthoryear {%
Nejjar%
, Zacharias%
, Stiehle%
\BCBL {}\ \BBA {} Weber%
}{%
Nejjar%
\ \protect \BOthers {.}}{%
{\protect \APACyear {2025}}%
}]{%
nejjar2025llms}
\APACinsertmetastar {%
nejjar2025llms}%
\begin{APACrefauthors}%
Nejjar, M.%
, Zacharias, L.%
, Stiehle, F.%
\BCBL {} Weber, I.%
\end{APACrefauthors}%
\unskip\
\newblock
\APACrefYearMonthDay{2025}{}{}.
\newblock
{\BBOQ}\APACrefatitle {LLMs for science: Usage for code generation and data analysis} {Llms for science: Usage for code generation and data analysis}.{\BBCQ}
\newblock
\APACjournalVolNumPages{J. Softw. Evol. Process.}{37}{1}{,}
\newblock
\begin{APACrefDOI} \doi{10.1002/SMR.2723} \end{APACrefDOI}
\newblock

\newblock

\PrintBackRefs{\CurrentBib}

\bibitem [\protect \citeauthoryear {%
{Ollama}%
}{%
{Ollama}%
}{%
{\protect \APACyear {2024}}%
}]{%
ollama}
\APACinsertmetastar {%
ollama}%
\begin{APACrefauthors}%
{Ollama}%
\end{APACrefauthors}%
\unskip\
\newblock
\APACrefYearMonthDay{2024}{}{}.
\newblock
\APACrefbtitle {{O}llama.} {{O}llama.}
\newblock
\begin{APACrefURL} {https://ollama.com/} \end{APACrefURL}
\newblock
\APACrefnote{[Accessed 26-10-2024]}
\PrintBackRefs{\CurrentBib}

\bibitem [\protect \citeauthoryear {%
{OpenAI}%
}{%
{OpenAI}%
}{%
{\protect \APACyear {2024}}%
}]{%
openAI}
\APACinsertmetastar {%
openAI}%
\begin{APACrefauthors}%
{OpenAI}%
\end{APACrefauthors}%
\unskip\
\newblock
\APACrefYearMonthDay{2024}{}{}.
\newblock
\APACrefbtitle {OpenAI Codex.} {Openai codex.}
\newblock
\APAChowpublished {\url{https://openai.com/index/openai-codex/}}.
\newblock
\APACrefnote{[Accessed 03-11-2024]}
\PrintBackRefs{\CurrentBib}

\bibitem [\protect \citeauthoryear {%
Phung%
\ \protect \BOthers {.}}{%
Phung%
\ \protect \BOthers {.}}{%
{\protect \APACyear {2023}}%
}]{%
phung2023generatinghighprecisionfeedbackprogramming}
\APACinsertmetastar {%
phung2023generatinghighprecisionfeedbackprogramming}%
\begin{APACrefauthors}%
Phung, T.%
, Cambronero, J.%
, Gulwani, S.%
, Kohn, T.%
, Majumdar, R.%
, Singla, A.%
\BCBL {} Soares, G.%
\end{APACrefauthors}%
\unskip\
\newblock
\APACrefYearMonthDay{2023}{}{}.
\newblock
\APACrefbtitle {Generating High-Precision Feedback for Programming Syntax Errors using Large Language Models.} {Generating high-precision feedback for programming syntax errors using large language models.}
\newblock
\begin{APACrefURL} {https://arxiv.org/abs/2302.04662} \end{APACrefURL}
\PrintBackRefs{\CurrentBib}

\bibitem [\protect \citeauthoryear {%
Prather%
\ \protect \BOthers {.}}{%
Prather%
\ \protect \BOthers {.}}{%
{\protect \APACyear {2023}}%
}]{%
10.1145/3587103.3594206}
\APACinsertmetastar {%
10.1145/3587103.3594206}%
\begin{APACrefauthors}%
Prather, J.%
, Denny, P.%
, Leinonen, J.%
, Becker, B.A.%
, Albluwi, I.%
, Caspersen, M.E.%
\BDBL {}Savelka, J.%
\end{APACrefauthors}%
\unskip\
\newblock
\APACrefYearMonthDay{2023}{}{}.
\newblock
{\BBOQ}\APACrefatitle {Transformed by Transformers: Navigating the AI Coding Revolution for Computing Education: An ITiCSE Working Group Conducted by Humans} {Transformed by transformers: Navigating the ai coding revolution for computing education: An iticse working group conducted by humans}.{\BBCQ}
\newblock
 \APACrefbtitle {Proceedings of the 2023 Conference on Innovation and Technology in Computer Science Education V. 2} {Proceedings of the 2023 conference on innovation and technology in computer science education v. 2}\ (\BPG~561–562).
\newblock
\APACaddressPublisher{New York, NY, USA}{Association for Computing Machinery}.
\newblock
\begin{APACrefURL} {https://doi.org/10.1145/3587103.3594206} \end{APACrefURL}
\PrintBackRefs{\CurrentBib}

\bibitem [\protect \citeauthoryear {%
Prather%
, Denny%
\BCBL {}\ \protect \BOthers {.}}{%
Prather%
, Denny%
\BCBL {}\ \protect \BOthers {.}}{%
{\protect \APACyear {2023}}%
}]{%
10.1145/3623762.3633499}
\APACinsertmetastar {%
10.1145/3623762.3633499}%
\begin{APACrefauthors}%
Prather, J.%
, Denny, P.%
, Leinonen, J.%
, Becker, B.A.%
, Albluwi, I.%
, Craig, M.%
\BDBL {}Savelka, J.%
\end{APACrefauthors}%
\unskip\
\newblock
\APACrefYearMonthDay{2023}{}{}.
\newblock
{\BBOQ}\APACrefatitle {The Robots Are Here: Navigating the Generative AI Revolution in Computing Education} {The robots are here: Navigating the generative ai revolution in computing education}.{\BBCQ}
\newblock
 \APACrefbtitle {Proceedings of the 2023 Working Group Reports on Innovation and Technology in Computer Science Education} {Proceedings of the 2023 working group reports on innovation and technology in computer science education}\ (\BPG~108–159).
\newblock
\APACaddressPublisher{New York, NY, USA}{Association for Computing Machinery}.
\newblock
\begin{APACrefURL} {https://doi.org/10.1145/3623762.3633499} \end{APACrefURL}
\PrintBackRefs{\CurrentBib}

\bibitem [\protect \citeauthoryear {%
Prather%
, Reeves%
\BCBL {}\ \protect \BOthers {.}}{%
Prather%
, Reeves%
\BCBL {}\ \protect \BOthers {.}}{%
{\protect \APACyear {2023}}%
}]{%
Prather_2023}
\APACinsertmetastar {%
Prather_2023}%
\begin{APACrefauthors}%
Prather, J.%
, Reeves, B.N.%
, Denny, P.%
, Becker, B.A.%
, Leinonen, J.%
, Luxton-Reilly, A.%
\BDBL {}Santos, E.A.%
\end{APACrefauthors}%
\unskip\
\newblock
\APACrefYearMonthDay{2023}{}{}.
\newblock
{\BBOQ}\APACrefatitle {{“It’s Weird That it Knows What I Want”}: Usability and Interactions with {Copilot} for Novice Programmers} {{“It’s Weird That it Knows What I Want”}: Usability and interactions with {Copilot} for novice programmers}.{\BBCQ}
\newblock
\APACjournalVolNumPages{ACM Transactions on Computer-Human Interaction}{31}{1}{1–31,}
\newblock
\begin{APACrefDOI} \doi{10.1145/3617367} \end{APACrefDOI}
\newblock

\newblock

\PrintBackRefs{\CurrentBib}

\bibitem [\protect \citeauthoryear {%
Price%
\ \protect \BOthers {.}}{%
Price%
\ \protect \BOthers {.}}{%
{\protect \APACyear {2019}}%
}]{%
price2019comparison}
\APACinsertmetastar {%
price2019comparison}%
\begin{APACrefauthors}%
Price, T.W.%
, Dong, Y.%
, Zhi, R.%
, Paa{\ss}en, B.%
, Lytle, N.%
, Catet{\'e}, V.%
\BCBL {} Barnes, T.%
\end{APACrefauthors}%
\unskip\
\newblock
\APACrefYearMonthDay{2019}{}{}.
\newblock
{\BBOQ}\APACrefatitle {A comparison of the quality of data-driven programming hint generation algorithms} {A comparison of the quality of data-driven programming hint generation algorithms}.{\BBCQ}
\newblock
\APACjournalVolNumPages{International Journal of Artificial Intelligence in Education}{29}{}{368--395,}
\newblock
\begin{APACrefDOI} \doi{10.1007/s40593-019-00177-z} \end{APACrefDOI}
\newblock

\newblock

\PrintBackRefs{\CurrentBib}

\bibitem [\protect \citeauthoryear {%
Qwen%
}{%
Qwen%
}{%
{\protect \APACyear {2024}}%
}]{%
huggingfaceQwenCodeQwen157BHugging}
\APACinsertmetastar {%
huggingfaceQwenCodeQwen157BHugging}%
\begin{APACrefauthors}%
Qwen%
\end{APACrefauthors}%
\unskip\
\newblock
\APACrefYearMonthDay{2024}{}{}.
\newblock
\APACrefbtitle {{Q}wen/{C}ode{Q}wen1.5-7{B} · {H}ugging {F}ace --- huggingface.co.} {{Q}wen/{C}ode{Q}wen1.5-7{B} · {H}ugging {F}ace --- huggingface.co.}
\newblock
\APAChowpublished {\url{https://huggingface.co/Qwen/CodeQwen1.5-7B}}.
\newblock
\APACrefnote{[Accessed 18-02-2025]}
\PrintBackRefs{\CurrentBib}

\bibitem [\protect \citeauthoryear {%
Raschka%
, Patterson%
\BCBL {}\ \BBA {} Nolet%
}{%
Raschka%
\ \protect \BOthers {.}}{%
{\protect \APACyear {2020}}%
}]{%
info11040193}
\APACinsertmetastar {%
info11040193}%
\begin{APACrefauthors}%
Raschka, S.%
, Patterson, J.%
\BCBL {} Nolet, C.%
\end{APACrefauthors}%
\unskip\
\newblock
\APACrefYearMonthDay{2020}{}{}.
\newblock
{\BBOQ}\APACrefatitle {Machine Learning in Python: Main Developments and Technology Trends in Data Science, Machine Learning, and Artificial Intelligence} {Machine learning in python: Main developments and technology trends in data science, machine learning, and artificial intelligence}.{\BBCQ}
\newblock
\APACjournalVolNumPages{Information}{11}{4}{,}
\newblock
\begin{APACrefDOI} \doi{10.3390/info11040193} \end{APACrefDOI}
\newblock

\newblock

\PrintBackRefs{\CurrentBib}

\bibitem [\protect \citeauthoryear {%
Ribeiro%
}{%
Ribeiro%
}{%
{\protect \APACyear {2023}}%
}]{%
10.1145/3618305.3623587}
\APACinsertmetastar {%
10.1145/3618305.3623587}%
\begin{APACrefauthors}%
Ribeiro, F.%
\end{APACrefauthors}%
\unskip\
\newblock
\APACrefYearMonthDay{2023}{}{}.
\newblock
{\BBOQ}\APACrefatitle {Large Language Models for Automated Program Repair} {Large language models for automated program repair}.{\BBCQ}
\newblock
 \APACrefbtitle {Companion Proceedings of the 2023 ACM SIGPLAN International Conference on Systems, Programming, Languages, and Applications: Software for Humanity} {Companion proceedings of the 2023 acm sigplan international conference on systems, programming, languages, and applications: Software for humanity}\ (\BPG~7–9).
\newblock
\APACaddressPublisher{New York, NY, USA}{Association for Computing Machinery}.
\newblock
\begin{APACrefURL} {https://doi.org/10.1145/3618305.3623587} \end{APACrefURL}
\PrintBackRefs{\CurrentBib}

\bibitem [\protect \citeauthoryear {%
Skripchuk%
, Shi%
\BCBL {}\ \BBA {} Price%
}{%
Skripchuk%
\ \protect \BOthers {.}}{%
{\protect \APACyear {2022}}%
}]{%
skripchuk2022identifying}
\APACinsertmetastar {%
skripchuk2022identifying}%
\begin{APACrefauthors}%
Skripchuk, J.%
, Shi, Y.%
\BCBL {} Price, T.%
\end{APACrefauthors}%
\unskip\
\newblock
\APACrefYearMonthDay{2022}{}{}.
\newblock
{\BBOQ}\APACrefatitle {Identifying Common Errors in Open-Ended Machine Learning Projects} {Identifying common errors in open-ended machine learning projects}.{\BBCQ}
\newblock
 \APACrefbtitle {Proceedings of the 53rd ACM Technical Symposium on Computer Science Education - Volume 1} {Proceedings of the 53rd acm technical symposium on computer science education - volume 1}\ (\BPG~216–222).
\newblock
\APACaddressPublisher{New York, NY, USA}{Association for Computing Machinery}.
\newblock
\begin{APACrefURL} {https://doi.org/10.1145/3478431.3499397} \end{APACrefURL}
\PrintBackRefs{\CurrentBib}

\bibitem [\protect \citeauthoryear {%
S~Kumar%
, Adam~Lones%
, Maarek%
\BCBL {}\ \BBA {} Zantout%
}{%
S~Kumar%
\ \protect \BOthers {.}}{%
{\protect \APACyear {2024}}%
}]{%
10.1145/3643795.3648380}
\APACinsertmetastar {%
10.1145/3643795.3648380}%
\begin{APACrefauthors}%
S~Kumar, S.%
, Adam~Lones, M.%
, Maarek, M.%
\BCBL {} Zantout, H.%
\end{APACrefauthors}%
\unskip\
\newblock
\APACrefYearMonthDay{2024}{}{}.
\newblock
{\BBOQ}\APACrefatitle {Investigating the Proficiency of Large Language Models in Formative Feedback Generation for Student Programmers} {Investigating the proficiency of large language models in formative feedback generation for student programmers}.{\BBCQ}
\newblock
 \APACrefbtitle {Proceedings of the 1st International Workshop on Large Language Models for Code} {Proceedings of the 1st international workshop on large language models for code}\ (\BPG~88–93).
\newblock
\APACaddressPublisher{New York, NY, USA}{Association for Computing Machinery}.
\newblock
\begin{APACrefURL} {https://doi.org/10.1145/3643795.3648380} \end{APACrefURL}
\PrintBackRefs{\CurrentBib}

\bibitem [\protect \citeauthoryear {%
Sobania%
, Briesch%
, Hanna%
\BCBL {}\ \BBA {} Petke%
}{%
Sobania%
\ \protect \BOthers {.}}{%
{\protect \APACyear {2023}}%
}]{%
sobania2023analysisautomaticbugfixing}
\APACinsertmetastar {%
sobania2023analysisautomaticbugfixing}%
\begin{APACrefauthors}%
Sobania, D.%
, Briesch, M.%
, Hanna, C.%
\BCBL {} Petke, J.%
\end{APACrefauthors}%
\unskip\
\newblock
\APACrefYearMonthDay{2023}{}{}.
\newblock
\APACrefbtitle {An Analysis of the Automatic Bug Fixing Performance of ChatGPT.} {An analysis of the automatic bug fixing performance of chatgpt.}
\newblock
\begin{APACrefURL} {https://arxiv.org/abs/2301.08653} \end{APACrefURL}
\PrintBackRefs{\CurrentBib}

\bibitem [\protect \citeauthoryear {%
Sulmont%
, Patitsas%
\BCBL {}\ \BBA {} Cooperstock%
}{%
Sulmont%
\ \protect \BOthers {.}}{%
{\protect \APACyear {2019}}%
}]{%
sulmont2019can}
\APACinsertmetastar {%
sulmont2019can}%
\begin{APACrefauthors}%
Sulmont, E.%
, Patitsas, E.%
\BCBL {} Cooperstock, J.R.%
\end{APACrefauthors}%
\unskip\
\newblock
\APACrefYearMonthDay{2019}{}{}.
\newblock
{\BBOQ}\APACrefatitle {Can You Teach Me To Machine Learn?} {Can you teach me to machine learn?}{\BBCQ}
\newblock
 \APACrefbtitle {Proceedings of the 50th ACM Technical Symposium on Computer Science Education} {Proceedings of the 50th acm technical symposium on computer science education}\ (\BPG~948–954).
\newblock
\APACaddressPublisher{New York, NY, USA}{Association for Computing Machinery}.
\newblock
\begin{APACrefURL} {https://doi.org/10.1145/3287324.3287392} \end{APACrefURL}
\PrintBackRefs{\CurrentBib}

\bibitem [\protect \citeauthoryear {%
Tambon%
\ \protect \BOthers {.}}{%
Tambon%
\ \protect \BOthers {.}}{%
{\protect \APACyear {2025}}%
}]{%
tambon2024bugs}
\APACinsertmetastar {%
tambon2024bugs}%
\begin{APACrefauthors}%
Tambon, F.%
, Dakhel, A.M.%
, Nikanjam, A.%
, Khomh, F.%
, Desmarais, M.C.%
\BCBL {} Antoniol, G.%
\end{APACrefauthors}%
\unskip\
\newblock
\APACrefYearMonthDay{2025}{}{}.
\newblock
{\BBOQ}\APACrefatitle {Bugs in large language models generated code: an empirical study} {Bugs in large language models generated code: an empirical study}.{\BBCQ}
\newblock
\APACjournalVolNumPages{Empir. Softw. Eng.}{30}{3}{65,}
\newblock
\begin{APACrefDOI} \doi{10.1007/S10664-025-10614-4} \end{APACrefDOI}
\newblock

\newblock

\PrintBackRefs{\CurrentBib}

\bibitem [\protect \citeauthoryear {%
Tu%
, Zou%
, Su%
\BCBL {}\ \BBA {} Zhang%
}{%
Tu%
\ \protect \BOthers {.}}{%
{\protect \APACyear {2024}}%
}]{%
tu2023should}
\APACinsertmetastar {%
tu2023should}%
\begin{APACrefauthors}%
Tu, X.%
, Zou, J.%
, Su, W.%
\BCBL {} Zhang, L.%
\end{APACrefauthors}%
\unskip\
\newblock
\APACrefYearMonthDay{2024}{{\APACmonth{01}}}{}.
\newblock
{\BBOQ}\APACrefatitle {What {Should} {Data} {Science} {Education} {Do} {With} {Large} {Language} {Models}?} {What {Should} {Data} {Science} {Education} {Do} {With} {Large} {Language} {Models}?}{\BBCQ}
\newblock
\APACjournalVolNumPages{Harvard Data Science Review}{6}{1}{,}
\newblock
\begin{APACrefDOI} \doi{10.1162/99608f92.bff007ab} \end{APACrefDOI}
\newblock
\APACrefnote{Publisher: The MIT Press}
\newblock

\newblock

\PrintBackRefs{\CurrentBib}

\bibitem [\protect \citeauthoryear {%
Weizenbaum%
}{%
Weizenbaum%
}{%
{\protect \APACyear {1966}}%
}]{%
weizenbaum1966eliza}
\APACinsertmetastar {%
weizenbaum1966eliza}%
\begin{APACrefauthors}%
Weizenbaum, J.%
\end{APACrefauthors}%
\unskip\
\newblock
\APACrefYearMonthDay{1966}{}{}.
\newblock
{\BBOQ}\APACrefatitle {ELIZA—a computer program for the study of natural language communication between man and machine} {Eliza—a computer program for the study of natural language communication between man and machine}.{\BBCQ}
\newblock
\APACjournalVolNumPages{Communications of the ACM}{9}{1}{36--45,}
\newblock
\begin{APACrefDOI} \doi{10.1145/365153.365168} \end{APACrefDOI}
\newblock

\newblock

\PrintBackRefs{\CurrentBib}

\bibitem [\protect \citeauthoryear {%
Wermelinger%
}{%
Wermelinger%
}{%
{\protect \APACyear {2023}}%
}]{%
wermelinger2023using}
\APACinsertmetastar {%
wermelinger2023using}%
\begin{APACrefauthors}%
Wermelinger, M.%
\end{APACrefauthors}%
\unskip\
\newblock
\APACrefYearMonthDay{2023}{}{}.
\newblock
{\BBOQ}\APACrefatitle {Using GitHub Copilot to solve simple programming problems} {Using github copilot to solve simple programming problems}.{\BBCQ}
\newblock
 \APACrefbtitle {Proceedings of the 54th ACM Technical Symposium on Computer Science Education V. 1} {Proceedings of the 54th acm technical symposium on computer science education v. 1}\ (\BPGS\ 172--178).
\newblock
\begin{APACrefURL} {https://doi.org/10.1145/3545945.3569830} \end{APACrefURL}
\PrintBackRefs{\CurrentBib}

\bibitem [\protect \citeauthoryear {%
Xia%
, Wei%
\BCBL {}\ \BBA {} Zhang%
}{%
Xia%
\ \protect \BOthers {.}}{%
{\protect \APACyear {2023}}%
}]{%
xia2023automated}
\APACinsertmetastar {%
xia2023automated}%
\begin{APACrefauthors}%
Xia, C.S.%
, Wei, Y.%
\BCBL {} Zhang, L.%
\end{APACrefauthors}%
\unskip\
\newblock
\APACrefYearMonthDay{2023}{}{}.
\newblock
{\BBOQ}\APACrefatitle {Automated Program Repair in the Era of Large Pre-Trained Language Models} {Automated program repair in the era of large pre-trained language models}.{\BBCQ}
\newblock
 \APACrefbtitle {Proceedings of the 45th International Conference on Software Engineering} {Proceedings of the 45th international conference on software engineering}\ (\BPG~1482–1494).
\newblock
\APACaddressPublisher{}{IEEE Press}.
\newblock
\begin{APACrefURL} {https://doi.org/10.1109/ICSE48619.2023.00129} \end{APACrefURL}
\PrintBackRefs{\CurrentBib}

\bibitem [\protect \citeauthoryear {%
Yi%
, Ahmed%
, Karkare%
, Tan%
\BCBL {}\ \BBA {} Roychoudhury%
}{%
Yi%
\ \protect \BOthers {.}}{%
{\protect \APACyear {2017}}%
}]{%
yi2017feasibility}
\APACinsertmetastar {%
yi2017feasibility}%
\begin{APACrefauthors}%
Yi, J.%
, Ahmed, U.Z.%
, Karkare, A.%
, Tan, S.H.%
\BCBL {} Roychoudhury, A.%
\end{APACrefauthors}%
\unskip\
\newblock
\APACrefYearMonthDay{2017}{}{}.
\newblock
{\BBOQ}\APACrefatitle {A feasibility study of using automated program repair for introductory programming assignments} {A feasibility study of using automated program repair for introductory programming assignments}.{\BBCQ}
\newblock
 \APACrefbtitle {Proceedings of the 2017 11th Joint Meeting on Foundations of Software Engineering} {Proceedings of the 2017 11th joint meeting on foundations of software engineering}\ (\BPGS\ 740--751).
\newblock
\begin{APACrefURL} {https://doi.org/10.1145/3106237.3106262} \end{APACrefURL}
\PrintBackRefs{\CurrentBib}

\bibitem [\protect \citeauthoryear {%
Zhang%
\ \protect \BOthers {.}}{%
Zhang%
\ \protect \BOthers {.}}{%
{\protect \APACyear {2022}}%
}]{%
zhang2022repairingbugspythonassignments}
\APACinsertmetastar {%
zhang2022repairingbugspythonassignments}%
\begin{APACrefauthors}%
Zhang, J.%
, Cambronero, J.%
, Gulwani, S.%
, Le, V.%
, Piskac, R.%
, Soares, G.%
\BCBL {} Verbruggen, G.%
\end{APACrefauthors}%
\unskip\
\newblock
\APACrefYearMonthDay{2022}{}{}.
\newblock
\APACrefbtitle {Repairing Bugs in Python Assignments Using Large Language Models.} {Repairing bugs in python assignments using large language models.}
\newblock
\begin{APACrefURL} {https://arxiv.org/abs/2209.14876} \end{APACrefURL}
\PrintBackRefs{\CurrentBib}

\bibitem [\protect \citeauthoryear {%
Zhao%
\ \protect \BOthers {.}}{%
Zhao%
\ \protect \BOthers {.}}{%
{\protect \APACyear {2024}}%
}]{%
zhao2024surveylargelanguagemodels}
\APACinsertmetastar {%
zhao2024surveylargelanguagemodels}%
\begin{APACrefauthors}%
Zhao, W.X.%
, Zhou, K.%
, Li, J.%
, Tang, T.%
, Wang, X.%
, Hou, Y.%
\BDBL {}Wen, J\BHBI R.%
\end{APACrefauthors}%
\unskip\
\newblock
\APACrefYearMonthDay{2024}{}{}.
\newblock
\APACrefbtitle {A Survey of Large Language Models.} {A survey of large language models.}
\newblock
\begin{APACrefURL} {https://arxiv.org/abs/2303.18223} \end{APACrefURL}
\PrintBackRefs{\CurrentBib}

\bibitem [\protect \citeauthoryear {%
Zimmermann%
, Allin%
\BCBL {}\ \BBA {} Zhang%
}{%
Zimmermann%
\ \protect \BOthers {.}}{%
{\protect \APACyear {2024}}%
}]{%
zimmermann2023common}
\APACinsertmetastar {%
zimmermann2023common}%
\begin{APACrefauthors}%
Zimmermann, R.M.%
, Allin, S.%
\BCBL {} Zhang, L.%
\end{APACrefauthors}%
\unskip\
\newblock
\APACrefYearMonthDay{2024}{}{}.
\newblock
{\BBOQ}\APACrefatitle {Common Errors in Machine Learning Projects: A Second Look} {Common errors in machine learning projects: A second look}.{\BBCQ}
\newblock
 \APACrefbtitle {Proceedings of the 23rd Koli Calling International Conference on Computing Education Research.} {Proceedings of the 23rd koli calling international conference on computing education research.}
\newblock
\APACaddressPublisher{New York, NY, USA}{Association for Computing Machinery}.
\newblock
\begin{APACrefURL} {https://doi.org/10.1145/3631802.3631808} \end{APACrefURL}
\PrintBackRefs{\CurrentBib}

\end{thebibliography}

\end{document}